\documentclass[11pt,letterpaper]{article}
\pdfoutput=1
\usepackage[utf8]{inputenc}
\usepackage{mathtools}
\usepackage{axodraw2}
\usepackage{amscd}
\usepackage{amsmath}
\usepackage{slashed}
\usepackage{bm}
\usepackage{jheppub} 
\usepackage{caption} 
\usepackage{subcaption}

\title{Flavonstrahlung in the {\boldmath $B_3-L_2$ $Z^\prime$} Model at
  Current and Future Colliders}

\preprint{CERN-TH-2022-188}

\author[a,b]{Ben Allanach}
\affiliation[a]{DAMTP, University of Cambridge, Wilberforce Road, Cambridge, 
  CB3 0WA, United Kingdom}
\affiliation[b]{Department of Theoretical Physics, CERN, 1211 Geneva 23, Switzerland}
\emailAdd{B.C.Allanach@damtp.cam.ac.uk}
\author[a,1]{and Eetu Loisa\note{Corresponding author.}}
\emailAdd{eal47@cam.ac.uk}

\abstract{The $B_3-L_2$ $Z^\prime$ model may explain some gross features of
  the fermion mass spectrum as well as $b\rightarrow s \ell
  \ell$ anomalies. A TeV-scale physical scalar field associated with
  gauged $U(1)_{B_3-L_2}$ spontaneous symmetry breaking, the flavon field
  $\vartheta$, affects 
  Higgs phenomenology via mixing. In this paper, we investigate the 
  collider phenomenology of the flavon field. Higgs and $ W $ boson mass data are used to
  place bounds upon parameter space. We then examine
  \emph{flavonstrahlung} (${Z^\prime} \rightarrow Z^\prime \vartheta$
  production) at colliders as a means to directly produce and discover 
  flavon particles, which would provide
  direct empirical evidence tying the flavon to
  $U(1)_{B_3-L_2}$ symmetry breaking.
  A  100~TeV~FCC-hh or a 10~TeV muon collider would have high sensitivity to
  flavonstrahlung, whereas the HL-LHC can observe it only in extreme corners
  of parameter space. }

\newcommand{\ubl}{U(1)_{B_3-L_2}}
\newcommand{\bsmm}{$b \rightarrow s \mu^+ \mu^-$}
\newcommand{\p}{\partial} 
\newcommand{\hc}{^{\dagger}} 
\DeclareMathOperator{\diag}{diag} 
\let\originalleft\left
\let\originalright\right
\renewcommand{\left}{\mathopen{}\mathclose\bgroup\originalleft}
\renewcommand{\right}{\aftergroup\egroup\originalright}


\begin{document} 
\maketitle
\flushbottom

\section{Introduction \label{sec:intro}}
\begin{figure}
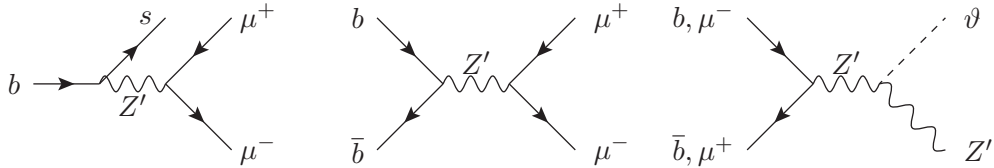

\begin{center}
  \begin{axopicture}(300,75)(-25,-5)
    \Line[arrow](-25,25)(0,50)
    \Line[arrow](-50,25)(-25,25)
    \Line[arrow](25,50)(0,25)
    \Line[arrow](0,25)(25,0)
    \Photon(-25,25)(0,25){3}{3}
    \Text(-12.5,17)[c]{$Z^{\prime}$}
    \Text(-55,25)[r]{$b$}
    \Text(-5,50)[r]{$s$}
    \Text(35,50)[c]{$\mu^+$}
    \Text(35,0)[c]{$\mu^-$}

    \Line[arrow](80,50)(105,25)
    \Line[arrow](105,25)(80,0)
    \Line[arrow](155,50)(130,25)
    \Line[arrow](130,25)(155,0)
    \Photon(105,25)(130,25){3}{3}
    \Text(117.5,33)[c]{$Z^\prime$}
    \Text(75,0)[r]{$\overline{b}$}
    \Text(75,50)[r]{$b$}
    \Text(162,50)[l]{$\mu^+$}
    \Text(162,0)[l]{$\mu^-$}

    \Line[arrow](220,50)(245,25)
    \Line[arrow](245,25)(220,0)
    \DashLine(295,50)(270,25){3}
    \Photon(270,25)(295,0){3}{3}
    \Photon(245,25)(270,25){3}{3}
    \Text(257.5,33)[c]{$Z^\prime$}
    \Text(215,0)[r]{$\overline{b}, \mu^+$}
    \Text(215,50)[r]{$b, \mu^-$}
    \Text(302,50)[l]{$\vartheta$}
    \Text(302,0)[l]{$Z^\prime$}
    \end{axopicture}
\end{center}
\caption{\label{fig:mmbar} Feynman diagram of (left panel)
$Z^\prime$ mediated contribution to $b \rightarrow
                           s \mu^+ \mu^-$, (middle panel) 
                           the dominant LHC $Z^\prime$
    production process, followed by subsequent decay into a di-muon pair
    $\mu^+ \mu^-$ and (right panel)  flavonstrahlung at a hadron collider
                           [$b \bar b$ initial state] or a muon collider
                           [$\mu^+ \mu^-$ initial state].} 
\end{figure}

The $B_3-L_2$ model~\cite{Alonso:2017uky,Bonilla:2017lsq,Allanach:2020kss} may be motivated by providing an explanation of some features of the
  fermion mass spectrum.
It was originally
introduced to explain discrepancies between 
Standard Model (SM) predictions and experimental measurements of various
observable quantities that involve the $b \rightarrow s \mu^+ \mu^-$ or
$\bar b \rightarrow \bar s \mu^+ \mu^-$
transition~\cite{Alguero:2019ptt,Alok:2019ufo,Ciuchini:2019usw,Aebischer:2019mlg,Datta:2019zca,Kowalska:2019ley,Arbey:2019duh,Gubernari:2022hxn,Alok:2022pjb,SinghChundawat:2022zdf}.\footnote{Discrepancies
between predictions and data are still present when one uses ratios of
observables to cancel the dependence upon CKM matrix elements,
whose determination from data is based on $\Delta M_{s,d}$,
$\epsilon_K$ and $S_{\psi K_S}$ for which we find 
new physics contributions that are
negligible~\cite{Buras:2022wpw,Buras:2022qip}.} Whilst four of these
  measurements (two different $q^2$ bins each of $R_K$ and $R_{K^\ast}$)
  have recently returned~\cite{LHCb:2022qnv,LHCb:2022zom} to
  being 
  compatible with SM predictions, a larger number of other observables are in
  tension 
  with them.
A $Z^\prime$ contribution to the $b \rightarrow s \mu^+ \mu^-$ process is
depicted in
the left-hand panel of figure~\ref{fig:mmbar}.  
The $B_3-L_2$ model is based on an extension of the SM gauge group by a direct
product with an 
additional spontaneously broken $U(1)_{B_3-L_2}$ gauge symmetry, where
the charges of the SM fields are proportional to third family baryon number
minus 
second family lepton number. The model is free of quantum field theoretic
anomalies if one includes
one right-handed neutrino Weyl fermion field per SM family in the field
content, a choice which 
is otherwise motivated by the fact that it facilitates the see-saw explanation of
the light neutrino masses inferred from empirical data. 
$U(1)_{B_3-L_2}$ is broken near the TeV scale by a \emph{flavon} field, a
SM-singlet complex scalar field $\theta$ with non-zero $B_3-L_2$ charge,
resulting in a TeV-scale electrically neutral gauge boson, i.e.\ a
$Z^\prime$ state. Much as the Higgs doublet field of
the SM possesses a physical state, the Higgs boson, which has been discovered
in experiments,
so the flavon field contains a physical real scalar state, the
flavon particle $\vartheta$.

It is remarkable that a model with spontaneous symmetry
breaking at a relatively low (i.e.\ TeV) scale can pass the experimental
bounds upon 
it coming from flavour constraints. Such constraints are notoriously strict
when they involve flavour transitions of the first two generations of
electrically charged fermionic fields, particularly from $K-\overline{K}$
mixing \cite{Charles:2013aka}.
Moreover, the strength of LHC bounds coming from bump hunts in the di-muon
mass spectrum resulting from $pp \rightarrow Z^\prime \rightarrow \mu^+ \mu^-$
is diminished by the fact that the $Z^\prime$ only
couples with an appreciable strength to \emph{third} family fermions. Thus,
the 
dominant LHC production process (as depicted in the middle panel of figure~\ref{fig:mmbar}) originates from
fusing a bottom quark $b$ and 
an anti-bottom quark $\bar b$ in the initial proton states, providing double
suppression to the $Z^\prime$ production cross section from small
(anti-)bottom parton distribution 
functions~\cite{Lim:2016wjo}. Current direct searches imply a lower bound upon
the $Z^\prime$ mass $M_{Z^\prime}$
of around 1-2 TeV~\cite{Bonilla:2017lsq,Allanach:2020kss,Allanach:2021gmj,Azatov:2022itm}, significantly lower than $Z^\prime$ models in which
the $Z^\prime$ field couples to quarks universally, where the current lower
bound from ATLAS and CMS is currently
around 5 TeV~\cite{ATLAS:2019erb,CMS:2021ctt}. A window in the parameter space
with 20~GeV $<M_{Z^\prime}<$ 300~GeV~\cite{Bonilla:2017lsq,Allanach:2020kss}
is all but ruled out at the 
$95\%$ confidence level~\cite{Allanach:2021gmj}. A recent
analysis~\cite{Azatov:2022itm} found that 
although through much of the parameter space the $Z^\prime$ can be
discovered at the HL-LHC, either a 3~TeV $\mu^+\mu^-$ 
collider, a
10~TeV
$\mu^+\mu^-$ 
collider~\cite{MuonCollider:2022nsa} or a 100~TeV FCC-hh 
collider 
\cite{FCC:2018vvp} would cover all of the parameter space compatible with
the model's
explanation of the $b\rightarrow s \mu^+\mu^-$
anomalies.  

It is remarkable that a TeV-scale $Z^\prime$, which generates flavour changing
neutral currents at the tree-level of perturbation theory, is not only allowed
by current data but in fact is motivated by it. Whether or not the
\bsmm\ anomalies persist in collider data, the $B_3-L_2$ model is of interest
both for 
this reason and because it may be pertinent to the fermion mass puzzle (namely
why certain hierarchies within the spectrum of fermion masses and mixings
exist). Aside 
from the aforementioned $Z^\prime$ production, scant collider
phenomenology of the model has been studied in terms of direct searches. 

It is our purpose here to study the direct collider phenomenology of the
flavon in the $B_3-L_2$ model for the first time. The flavon is 
expected to have a mass of
order $M_{Z^\prime}$ (i.e.\ the TeV scale) and so may be amenable to
direct production and discovery at high energy colliders. We shall see that
the flavon field
generically mixes with the Higgs field and is therefore bounded by some
electroweak measurements and Higgs searches. 
In detail, we wish to make a first estimate of the current and potential
future collider 
capabilities of observing the flavon via
flavonstrahlung~\cite{Allanach:2020kss}\footnote{LHC $HZ^\prime$ production
has been studied in a family universal $U(1)$ model in refs.~\cite{Pruna:2011me,Banerjee:2015hoa}.}, a process
whose dominant Feynman diagram is depicted
in the right-hand panel of figure~\ref{fig:mmbar}. Observation of a flavon
would be an important confirmation of the means of
$U(1)_{B_3-L_2}$ gauge
symmetry breakdown, since $U(1)_{B_3-L_2}$ breaking
could instead be a consequence of the Stueckelberg
mechanism~\cite{Kors:2005uz}, which 
possesses no explicit flavon field. This
then has implications for the fermion mass puzzle since, for example, the
Froggatt-Nielsen mechanism~\cite{Froggatt:1978nt}, which can explain the
hierarchies between the different families observed in
the measured values of the SM fermion masses, necessarily involves a flavon
field. 

The paper proceeds in section~\ref{sec:b3minusl2} by briefly reiterating some salient
points in the construction of
the $B_3-L_2$ model. The Higgs phenomenology of the flavon particle is reviewed
in section~\ref{sec:flavon:pheno} and
current collider constraints upon the parameter space of the $B_3-L_2$ model
are imposed. Then, in section~\ref{sec:prod}, we study the flavonstrahlung process
in allowed parts of the parameter space. We shall see that the cross-section
is too small to result in a realistic measurement at the HL-LHC (except for a
small portion of parameter space in the case that the flavon charge is larger
than unity) but that a 
10~TeV muon collider of a 100~TeV hadron collider such as the FCC-hh could
facilitate discovery of flavonstrahlung and therefore
facilitate discovery of the flavon field associated with breaking the
gauged flavour symmetry.
We summarise and conclude in section~\ref{sec:conc}.

\section{\boldmath{$B_3-L_2$} Model \label{sec:b3minusl2}}
We shall now review some salient points of the construction of the $B_3-L_2$
model. For definiteness, we use the simple bottom-up
construction of 
ref.~\cite{Allanach:2020kss}. 
The $ B_3 - L_2 $ model is constructed by extending the SM gauge group, $
SU(3) \times SU(2)_L \times U(1)_Y$, by an abelian factor $ \ubl $ in a
direct product. This new symmetry is gauged, so it comes with an
electrically neutral force carrier, the $ Z^\prime $ boson. We also introduce the
flavon field $ \left( \theta \right)  $, a SM-singlet complex scalar, which
carries charge $ q_{\theta} $ under $ \ubl $. Table~\ref{tab:charges} displays
the charges of the fields in the model under the new abelian symmetry. Three
right-handed neutrinos $\nu_{R_{1,2,3}}$ are added in order to facilitate
neutrino masses and 
mixing. With these fields and charges, $ \ubl $ acts vectorially on the
fermions (i.e.\ acts in the same way on left-handed chiral fermions
as it does on their right-handed partners), and gauge anomaly cancellation
manifestly takes place. There is an implicit assumption in the $B_3-L_2$ model
(and other similar models) that
they originate from some more complete ultra-violet model which may be a
semi-simple extension, thus obviating constraints coming from 
Landau poles~\cite{Bause:2021prv}.

\begin{table}[htpb]
	\centering

	\begin{tabular}{ccccccccc}
		\hline \hline
		$ Q'_{iL} $ & $ u'_{iR} $ & $ d'_{iR} $ & $ L_1' $& $ L_2' $ & $ L_3' $ & $ e'_{1 R} $ & $ e'_{2R} $ & $ e'_{3R} $ \\
		0 & 0 & 0 & 0 & -3 & 0 & 0 & -3 & 0  \\
		\hline
		$\nu'_{1R} $ & $\nu'_{2R} $ & $\nu'_{3R} $ & $ Q'_{3L} $ & $ u'_{3R} $ & $ d'_{3R} $ & $ H $ & $ \theta $ \\
	0 & -3 & 0 & 1 & 1 & 1 & 0 & $ q_{\theta} $ \\
	\hline \hline
	\end{tabular}
	\caption{The $\ubl$ charge assignments. A prime stands for a weak
          eigenstate Weyl fermion and the family index $ i $ takes values 1
          and 2. The flavon charge $ q_{\theta} $ is a non-zero rational
  number. \label{tab:charges}}
\end{table}

The fermionic couplings of the $ Z^\prime $ are expressed in the Lagrangian density
\begin{equation} \label{eq:zprime_fermion_lagr_weak}
	\mathcal{L}_{Z^\prime \psi} =
	- g_{Z^\prime} \left( 
		\overline{Q'_{3L}} \slashed{Z}^\prime Q'_{3L} 
	+ \overline{u'_{3R}} \slashed{Z}^\prime u'_{3R} 
+  \overline{d'_{3R}} \slashed{Z}^\prime d'_{3R}  
- 3 \overline{L'_{2L}} \slashed{Z}^\prime L_{2L}' 
- 3 \overline{e'_{2R}} \slashed{Z}^\prime e'_{2R}  
- 3 \overline{\nu'_{2R}} \slashed{Z}^\prime \nu'_{2R} \right) 
\end{equation}
where $ g_{Z^\prime} $ is the gauge coupling of the new $ \ubl $ symmetry. In order to study the phenomenology of the model, however, we must transform to the mass basis. We denote the 3-component column vectors in family space by boldface letters,  $ \mathbf{Q_L}' = \left( \mathbf{u_L}' , \mathbf{d_L}' \right)  $, $ \mathbf{L_L}' = \left(\mathbf{v_L}', \mathbf{e_L}'  \right)  $,  $ \mathbf{u_R}' $, $ \mathbf{d_R}' $,  $\mathbf{e_R}'$ and $ \boldsymbol{\nu}_\mathbf{R}' $. The transformation between the (primed) weak eigenbasis and the (unprimed) mass eigenbasis written
\begin{equation}
	\mathbf{P}' = V_I \mathbf{P}
\end{equation}
for $ I \in \left\{ u_L, d_L, e_L,\nu_L, u_R, d_R,e_R, \nu_R \right\}  $. Encoding the family-dependent couplings of eq.~\ref{eq:zprime_fermion_lagr_weak} into two 3 by 3 diagonal matrices
\begin{equation}
	\Xi := 
	\begin{pmatrix} 
		0 & 0 & 0 \\
		0 & 0 & 0 \\
		0 & 0 & 1 \\
	\end{pmatrix} 
	, \quad
	\Omega := 
	\begin{pmatrix} 
		0 & 0 & 0 \\
		0 & 1 & 0 \\
		0 & 0 & 0 \\
	\end{pmatrix} 
\end{equation}
and defining $ \Lambda := V_I\hc \alpha V_I $ where $ I \in  \left\{ u_L, d_L, e_L, \nu_L, u_R, d_R, e_R, \nu_R \right\}, \alpha \in \left\{ \Xi, \Omega \right\} 	  $, we obtain $\mathcal{L}_{Z^\prime \psi}$ in the mass eigenbasis:
\begin{multline} \label{eq:zprime_fermion_lagr_mass}
	\mathcal{L}_{Z^\prime \psi}	=
	- g_{Z^\prime} \Big( \overline{\mathbf{u_{L}}} \Lambda_{\Xi}^{\left( u_L \right) } \slashed{Z}^\prime \mathbf{u_{L}} +
		\overline{\mathbf{d_L}} \Lambda_{\Xi}^{\left( d_L \right) } \slashed{Z}^\prime \mathbf{d_{L}} +
		\overline{\mathbf{u_{R}}} \Lambda_{\Xi}^{\left( u_R \right) } \slashed{Z}^\prime \mathbf{u_{R}} +
		\overline{\mathbf{d_{R}}} \Lambda_{\Xi}^{\left( d_R \right) } \slashed{Z} \mathbf{d_{R}} \\
	- 3 \overline{\boldsymbol{\nu}_{\mathbf L}} \Lambda_{\Omega}^{(\nu_L)} \slashed{Z}^\prime \boldsymbol{\nu}_{\mathbf L} 
	- 3 \overline{\mathbf{e_{L}}} \Lambda_{\Omega}^{(e_L)} \slashed{Z}^\prime \mathbf{e_{L}} 
- 3 \overline{\boldsymbol{\nu}_{\mathbf R}} \Lambda_{\Omega}^{\left( \nu_R \right) } \slashed{Z}^\prime \boldsymbol{\nu}_{\mathbf R} 
	- 3 \overline{\mathbf{e_{R}}} \Lambda_{\Omega}^{\left( e_R \right) } \slashed{Z}^\prime \mathbf{e_{R}}  \Big).
\end{multline}
Provided that $ \left(V_{d_L} \right)_{23} \neq 0 $,
eq.~\ref{eq:zprime_fermion_lagr_mass} couples the $ Z^\prime $ to both
$ (\overline{b} s + \overline{s} b) $ and $ \mu^+ \mu^- $. The $Z^\prime$
boson can thus mediate \bsmm transitions, thereby influencing $ B
$-observables.    

The kinetic term of the flavon field reads
\begin{equation} \label{eq:flavon_kin}
	\mathcal{L}_{\theta, \text{kin}} = \left( D^{\mu} \theta \right)^\ast \left( D_{\mu} \theta \right), 
\end{equation}
with its covariant derivative being
\begin{equation}
	D_{\mu} \theta = \left( \p _{\mu} - i g_{Z^\prime} q_{\theta} Z^\prime_\mu \right) \theta.
\end{equation}
The flavon field spontaneously breaks $ \ubl $ by developing a non-zero vacuum
expectation value 
(VEV) $ \langle \theta \rangle = v_{\theta} \neq 0 $. As a consequence, the $
Z^\prime $ acquires a mass $ M_{Z^\prime} = q_{\theta} g_{Z^\prime} v_{\theta} $. In the unitary
gauge, the $ Z^\prime $ boson eats the massless Goldstone boson associated with the
broken symmetry and obtains a longitudinal polarisation mode. We are left with
three $ Z^\prime $ degrees of freedom and a single real flavon field, $\vartheta$.

\subsection{Spontaneous symmetry breaking}\label{sec:flavon_potential}
 The introduction of a new SM-singlet scalar field, real or complex, modifies
 the scalar potential of the theory. In addition to the mass and quartic
 self-interaction terms for the new scalar, the theory also allows for a
 renormalisable dimension-4 term connecting the Higgs doublet to the
 beyond the SM
 (BSM)
 scalar. Thus, the combined scalar potential for the SM-Higgs doublet $ (H) $
 and the flavon field $ (\vartheta) $ in the $ B_3 - L_2 $ model reads
\begin{equation} \label{eq:scalar_potential}
	V(H, \theta) = -\mu_{H}^2 H^{\dagger} H + \lambda_H (H \hc H )^2 - \mu_{\theta}^2 \theta^\ast \theta + \lambda_{\theta} (\theta^\ast \theta )^2 + 	\lambda_{\theta H} \theta^\ast \theta H^{\dagger} H.
\end{equation}
Scalar potentials of this form have received considerable attention in the past (see for instance \cite{Barger:2008jx,Pruna:2013bma,Falkowski:2015iwa}), and we review here some standard steps. To find the lowest energy state of the potential, we work in the unitary gauge and expand both fields about their vacuum expectation values (VEVs), denoted by $ v_H $ and $ v_{\theta}$:
\begin{equation} 
	H = 
	\begin{pmatrix}
		0 \\
		\frac{v_H + h'}{\sqrt{2}}  \\
	\end{pmatrix}, 
	\qquad \theta = \frac{v_{\theta}  + \vartheta'}{\sqrt{2}}.
\end{equation}
Minimising the scalar potential with respect to the VEVs,
\begin{equation}
	\frac{\p V}{\p v_H} = 0, \quad \frac{\p V}{\p v_{\theta}} = 0,
\end{equation}
gives us the following two expressions:
\begin{align}
	v_H &= \sqrt{
	\frac{ 4 \mu_H^2 \lambda_{\theta} - 2 \mu_{\theta}^2 \lambda_{\theta H} }{ 4 \lambda_H \lambda_{\theta} - \lambda_{\theta H}^2 } \label{eq:higgsvev}
	}, \\
	v_{\theta} &= \sqrt{
	\frac{ 4 \mu_{\theta}^2 \lambda_{H} - 2 \mu_{H}^2 \lambda_{\theta H} }{ 4 \lambda_H \lambda_{\theta} - \lambda_{\theta H}^2 } \label{eq:flavonvev}
	}.
\end{align}
The requirements that the extremum be a local minimum and the potential
bounded from below for large field values provide the constraints 
\begin{align}
	4 \lambda_H \lambda_{\theta} - \lambda_{\theta H}^2 &> 0, \label{eq:condition_1}\\
	\lambda_H, \lambda_{\theta} > 0, \label{eq:condition_2}
\end{align}
respectively \cite{Pruna:2013bma}.

When we substitute the VEVs and the expanded scalar fields into eq.~\ref{eq:scalar_potential}, terms bilinear in $ h' $ and $ \vartheta' $ appear. Writing the quadratic part of the potential as 
\begin{equation}
	\mathcal{L}_{\text{quadratic}} = - \frac{1}{2} 
	\begin{pmatrix} 
		h' &
		\vartheta' \\
	\end{pmatrix}
	 M^2
	\begin{pmatrix} 
		h' \\
		\vartheta' \\
	\end{pmatrix}	
\end{equation}
leads to the non-diagonal, symmetric mass matrix
\begin{align}
	M^2 &= \begin{pmatrix}
 		\frac{1}{2} \lambda _{\theta H} v_{\theta }^2 + 3 \lambda_H v_H^2 - \mu _H^2 &  \lambda _{\theta H} v_H v_{\theta } \\
		  \lambda _{\theta H} v_H v_{\theta } & \frac{1}{2} \lambda _{\theta H} v_H^2 + 3 \lambda _{\theta } v_{\theta }^2 - \mu _{\theta }^2 \\	  
	\end{pmatrix}, \nonumber \\
	& = 
	\begin{pmatrix}
		2 \lambda_H v_H^2 & \lambda_{\theta H} v_H v_{\theta} \\
		\lambda_{\theta H} v_H v_{\theta} & 2 \lambda_{\theta} v_{\theta}^2 \\
	\end{pmatrix}.
\end{align}
The matrix $ M^2 $ is diagonalised by an orthogonal rotation $ P $, parameterised by an angle $ \phi $:  
\begin{align}
	M^2 &= P^T \diag\left( m_h^2, m_{\vartheta}^2 \right)  P ,\\
	\begin{pmatrix}
		2 \lambda_H v_H^2 & \lambda_{\theta H} v_H v_{\theta} \\
		\lambda_{\theta H} v_H v_{\theta} & 2 \lambda_{\theta} v_{\theta}^2 \\
	\end{pmatrix}
	&= 
	\begin{pmatrix}
		\cos \phi & \sin \phi \\
		-\sin \phi & \cos \phi \\
	\end{pmatrix}
	\begin{pmatrix}
		m_h^2 & 0 \\
		0 & m_{\vartheta}^2 \\
	\end{pmatrix}
	\begin{pmatrix}
		\cos \phi & -\sin \phi \\
		\sin \phi & \cos \phi \\
	\end{pmatrix}.
\end{align}
Solving for $ \phi$, we find
\begin{equation} \label{eq:sin2phi}
	\sin 2 \phi = \frac{2 \lambda_{\theta H} v_H v_{\theta}}{m_{\vartheta}^2 - m_{h}^2}
\end{equation}
or, via a double angle formula, 
\begin{equation}
	\sin \phi  =  \sqrt{\frac{1}{2} \left(1-\sqrt{1-\frac{4 \lambda_{\theta H}^2 v_H^2 v_{\theta}^2}{\left(m_h^2 - m_{\vartheta}^2 	\right)^2}}\right)}.
\end{equation}
We have thus obtained the field rotation $ P $, which transforms the primed scalar field basis $ \left( h',\vartheta' \right)  $ into the (unprimed) mass eigenbasis:  
\begin{equation} \label{eq:rotation}
    \begin{pmatrix}
    h \\
    \vartheta \\
    \end{pmatrix}
    = 
    \begin{pmatrix}
    \cos \phi & -\sin \phi \\
    \sin \phi & \cos \phi \\
    \end{pmatrix}
    \begin{pmatrix}
    h' \\
    \vartheta' \\
\end{pmatrix}.
\end{equation}
The smallest eigenvalue of $ M^2 $, $ m_h^2 $, and the associated eigenstate $ h
$ are taken to 
correspond to the 125 GeV Higgs boson discovered at the LHC experiments,
whereas the larger-mass eigenstate $ \vartheta $ is the
flavon boson whose mass is written as $ m_{\vartheta}  $. 

Expanding the scalar fields about their VEVs and rotating into the mass
eigenbasis as described above, the kinetic term $ \mathcal{L}_{\theta,
  \text{kin}} $ yields an interaction term 
\begin{equation}
	 \mathcal{L}_{\theta, \text{kin}} \supset \cos \phi \, g_{Z^\prime}^2 q_{\theta}^2  v_{\theta}  \vartheta Z^\prime_{\mu} {Z^\prime}^{\mu}.
\end{equation}
This term helps give rise to the flavonstrahlung process depicted in the right-hand panel of figure~\ref{fig:mmbar} and it will therefore play a key role in the present study. 

The Higgs-flavon axis of the model can be expressed in terms of three parameters: $ m_{\vartheta}, \phi $ and $ v_{\theta} $, the last of which can be decomposed as $ v_{\theta} = M_{Z^\prime} / ({q_{\theta} g_{Z^\prime}}) $. Accordingly, the quartic couplings $ \lambda_{H} $ and $ \lambda_{\theta} $, together with eq.~\ref{eq:sin2phi}, become
\begin{align}
	\lambda_H &= \frac{m_h^2 \cos^2 \phi + m_{\vartheta}^2 \sin^2 \phi}{2 v_H^2}, \label{eq:lamh} \\
	\lambda_{\theta} &= \frac{m_{\vartheta}^2 \cos^2 \phi + m_h^2 \sin^2 \phi}{2 v_{\theta}^2}, \label{eq:lamth} \\
	\lambda_{\theta H} &= \frac{\sin(2 \phi) \left( m_{\vartheta}^2 - m_h^2 \right)}{2 v_H v_{\theta}}.
\end{align}

\subsection{Assumptions}
We must specify the model further before we can study its phenomenology. In
particular, the $V_I$ mixing matrices deserve our attention. With simplicity, ease of passing
flavour bounds and
the ability to explain the neutral current \bsmm\ anomalies as guiding
principles, an example set of mixing 
matrices $ V_{I} $ was proposed in ref.~\cite{Allanach:2020kss}: 
\begin{equation}
V_{d_L} = \begin{pmatrix} 
	1 & 0 & 0 \\
	0 & \cos \theta_{sb} &  - \sin \theta_{sb}\\
	0 & \sin \theta_{sb}& \cos \theta_{sb} \\
	\end{pmatrix},
\end{equation}
$ V_{d_R} = 1, V_{e_R} = 1, V_{e_L}=1 $ and $ V_{u_R} = 1$, where here $1$
denotes the 3 by 3 identity matrix. These imply that $
V_{u_{L}} = V_{d_L} V\hc $ and $ V_{\nu_L} = U\hc $, where $ V $ and $ U $ are
the CKM and PMNS matrices, respectively. Here, we will adhere to the same set of mixing matrices while keeping in mind that this choice is just intended to provide an example case for further study. This set of fermion mixing matrices results in a Lagrangian containing the terms 
\begin{equation} \label{eq:b3l2_c9}
	\mathcal{L} \supset - g_{Z^\prime} \left[\left(  \frac{1}{2} \sin 2 \theta_{sb} \overline{s} \slashed{Z}^\prime P_L b + \text{H.c.}  \right) - 3 \overline{\mu} \slashed{Z}^\prime \mu \right] ,
\end{equation}
where $P_L$ is a left-handed spinor helicity projection operator.
Once the $ Z^\prime $ is integrated out, these terms yield a contribution to
the Wilson coefficient
$ \mathcal{C}_9 $ from
\begin{equation}
	{\mathcal H}_{\text{WET}} = \ldots + \left[ \mathcal{C}_9 \mathcal{N} (\bar b \gamma_\alpha P_L
    s) (\bar \mu \gamma^\alpha \mu) + \text{H.c.}\right], \label{eq:Lwet}
\end{equation}
in the weak effective theory
Hamiltonian, which can
significantly ameliorate the 
\bsmm\ anomalies~\cite{Alguero:2019ptt,Alok:2019ufo,Ciuchini:2019usw,Aebischer:2019mlg,Datta:2019zca,Kowalska:2019ley,Arbey:2019duh,Gubernari:2022hxn}.
In eq.~\ref{eq:Lwet}, $\alpha \in \{0,1,2,3\}$ is a space-time index and
\begin{equation}
	{\mathcal N}= \frac{4 G_F}{\sqrt{2}}  V_{tb} V_{ts}^* \frac{\alpha}{4 \pi}= 1/(36\text{~TeV})^2
\end{equation}
is a normalising constant. 

We shall set the flavon charge $ q_{\theta} $ equal to 1 unless stated otherwise. The flavonstrahlung cross section is proportional to $ q_{\theta}^2 $, so it is straightforward to extend most of our results to different values of the charge.

\section{Flavon Phenomenology \label{sec:flavon:pheno}} 
In this section we first review the phenomenological constraints on the $ \ubl
$ model obtained in earlier work. These limits apply to the parameter set $
\{g_{Z^\prime}, M_{Z^\prime}, \theta_{23} \} $, the three inputs which
influence the ability of the model to explain \bsmm\ anomalies. We then move
on to study the flavon sector of the theory, obtaining an upper bound on the
Higgs-flavon mixing angle $ \phi $ and discussing the leading flavon
decay channels. 

We have updated the \textsc{FeynRules} \cite{Alloul:2013bka} implementation of
the $ \ubl $ model from ref.~\cite{Allanach:2020kss} by adding the flavon
sector, which was previously neglected.\footnote{The
model file can be found in the ancillary information of the {\textsc{arXiv}} version of this work.} We have also used \textsc{FeynRules} to convert the model into UFO format \cite{Degrande:2011ua}.

\subsection{Fit to neutral current \bsmm\ anomalies and LHC constraints} \label{sec:fit_to_NCBA}
One may use eq.~\ref{eq:b3l2_c9} to match the $ \ubl $ model to fits of \bsmm
data, as was done in ref.~\cite{Allanach:2020kss}. This condition lets us
eliminate the mixing angle $ \theta_{sb}$ and leaves us with two free
parameters relevant to $B$ decay data: $ M_{Z^\prime} $ and $ g_{Z^\prime} $. 
Defining the dimensionless quantity $ x $ as 
\begin{equation}
	x \coloneqq g_{Z^\prime} \frac{1 \text{~TeV}}{M_{Z^\prime}},
\end{equation}
one obtains
\begin{equation}
\theta_{sb}=\frac{1}{2} \sin^{-1}\left(\frac{-5.1 \times 10^{-4}
  \mathcal{C}_9}{x^2}\right). \label{tsb}
\end{equation}
In this work, we use $ \mathcal{C}_9 = -0.73 \pm 0.15 $, which is the best-fit value obtained in \cite{Altmannshofer:2021qrr} prior to the recent LHCb updates \cite{LHCb:2022qnv, LHCb:2022zom} of the lepton flavour universality ratios $ R_{K} $ and $ R_{K^*} $. 

It was found in ref.~\cite{Allanach:2020kss} that there are both lower and
upper bounds on the value of $ x $. The lower bound stems from measurements of
$ B_s - \overline{B_s} $ mixing, which the $ Z^\prime$ contributes to at
tree-level. The upper bound originates from measurements of the neutrino
trident cross-section, $ \sigma \left(\nu_{\mu} N \rightarrow \nu_{\mu} N
\mu^+ \mu^- \right)$, which also receives $ Z^\prime $ contributions at
tree-level. We are left with the constraint $0.04 < x < 0.67$ \cite{Azatov:2022itm} which we will adhere to in this work. Substituting such values of $x$ into eq.~\ref{tsb}
implies that $\theta_{sb}$ is small. For small $\theta_{sb}$, the collider
phenomenology that we shall discuss is not sensitive to its precise value and
so should also be valid for up-to-date fits of $\mathcal{C}_9$ incorporating
the recent LHCb results.\footnote{See 
  \cite{Ciuchini:2022wbq} for a recent fit.} This is because
  flavonstrahlung production proceeds initially via $Z^\prime$ production,
  which dominantly proceeds via $b \bar b$ fusion and is proportional to
  $g_{Z^\prime}^2 \cos^4 \theta_{sb} \approx g_{Z^\prime}^2 [1 -
  2 \theta_{sb}^2 + \mathcal{O}(\theta_{sb}^4)]$. 

In ref.~\cite{Allanach:2021gmj}, a set of LHC constraints was recast on the $
M_{Z^\prime} - g_{Z^\prime} $ plane of the theory. Figure 10 of that paper shows that the previously allowed light $ Z' $ region of parameter space, with
$M_{Z^\prime}<0.3$~TeV, is all but ruled out, and in the rest of the paper, we
shall assume that $M_{Z^\prime}>1$~TeV. A recent recasting of the CMS high-mass di-lepton searches \cite{CMS:2021ctt} eliminated at 95\% CL
all parameter points with $ M_{Z^\prime} \lesssim 2 \text{~TeV}$ \cite{Azatov:2022itm}. 

\subsection{Perturbativity of $Z^\prime$ couplings}
Neglecting fermion massess in comparison with $ M_{Z^\prime} $, the partial decay rate of the $ Z^\prime $ turning into a pair of Weyl fermions reads
\begin{equation}
	\Gamma \left( Z^\prime \rightarrow f_i \bar{f}_i \right)  = \frac{C_i}{24 \pi } (q_i g_{Z^\prime})^2 M_{Z^\prime},
\end{equation}
where $ C_i $ is the number of colour degrees of freedoms of the fermion $ f_i
$, and $ q_i $ is its $ \ubl $ charge\footnote{Note that $\Gamma/M_{Z^\prime}$
is independent of rescaling all charges by absorbing the scaling in
$g_{Z^\prime}$.} in units of $ g_{Z^\prime} $, as assigned in
Table~\ref{tab:charges}. Summing over all fermion species (except for the
right-handed neutrinos which are assumed to be more massive than the $Z^\prime$) yields the equation
\begin{equation}
	\frac{\Gamma}{M_{Z^\prime}} = \frac{13 g_{Z^\prime}^2}{8 \pi}
\end{equation}
where $ \Gamma $ is the total width of the $ Z^\prime $. We impose the limit $ \Gamma / M_{Z^\prime} < 1 / 3 $ to ensure that our perturbative cross-section calculations remain valid. This translates into $ g_{Z^\prime} < \sqrt{8 \pi / 39}~=~0.80  $. 
The perturbativity condition together with a fit to \bsmm\  data lead to an
upper bound  
on the $ Z^\prime $ mass: $ M_{Z^\prime} < 20$~TeV.

\subsection{Constraints on the mixing angle $ \phi $}
\begin{figure}[htpb]
	\centering
	\includegraphics[width=0.8\textwidth]{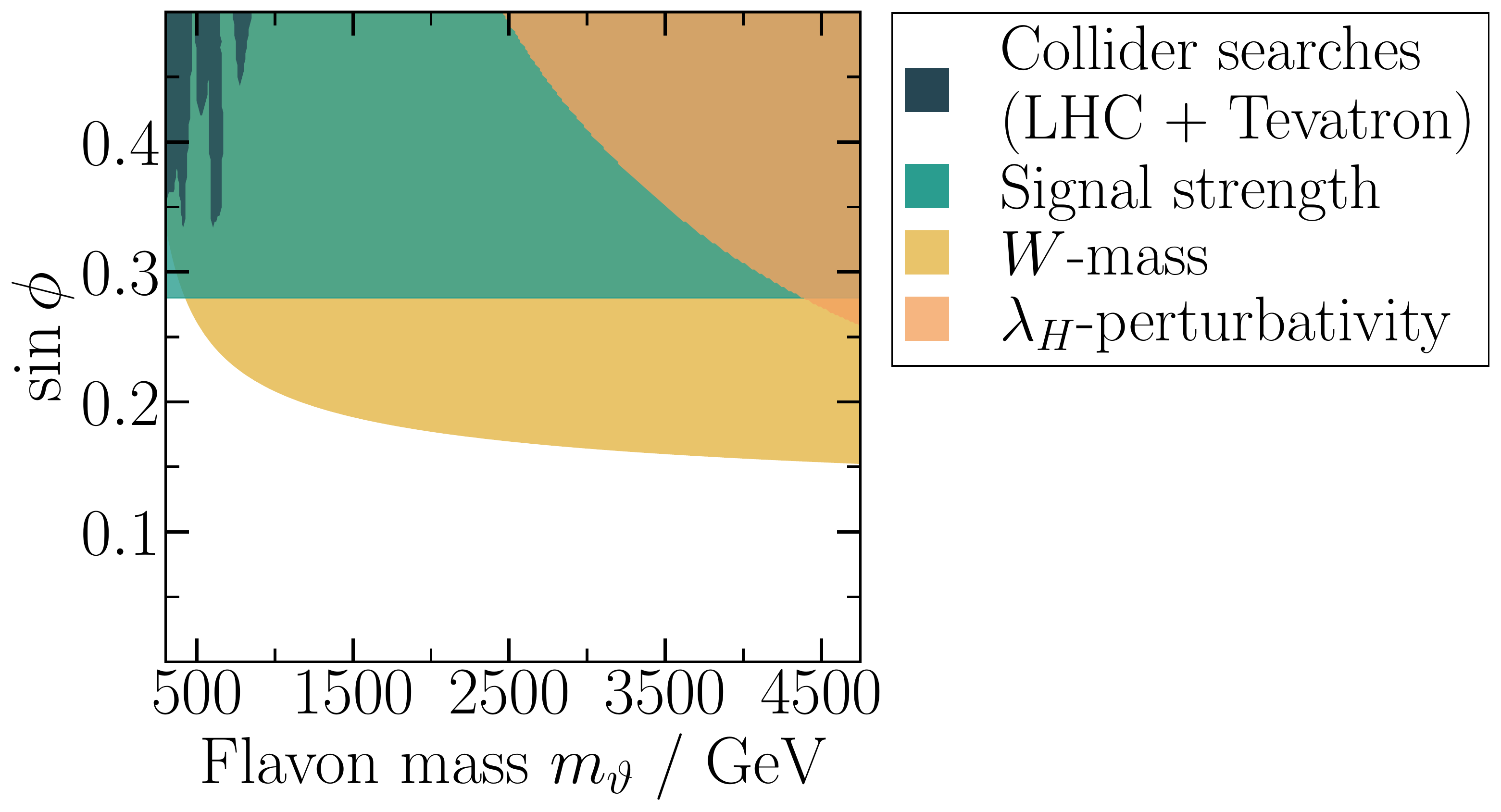}
	\caption{95\% CL limits on the Higgs-flavon mixing angle stemming from
	direct Higgs searches at the LHC and Tevatron (obtained
	using \textsc{HiggsBounds}), ATLAS signal strength measurements, the $ W $ boson mass and perturbativity of $ \lambda_H $. The white region is currently allowed. To obtain
	the $ M_{W} $-bound, we have assumed $ g_{Z^\prime} = 0.15, M_{Z^\prime}
	= 3$~TeV, but because $ M_W $ is only very weakly dependent on the
	$ Z^\prime $ parameters, we will use this bound for all values of $
	M_{Z^\prime} $ and $ g_{Z^\prime} $ considered in this work. \label{fig:mixing_exclusion}}
\end{figure}

There are numerous experimental and theoretical constraints on the
Higgs-flavon mixing angle $ \phi $. These are discussed at length in,
e.g.\ refs.~\cite{Robens:2015gla,Robens:2022cun, Falkowski:2015iwa} in the context of the
real singlet extension of the SM\@. We expect the constraints to be largely the
same in the $ B_3 - L_2 $ model, as most constraints are independent of the
presence of the $ Z^\prime $ and the extra degree of freedom from the
complexity of the flavon field. 
Because the $ B_3 - L_2 $ model is seen as a
low energy effective theory,
we disregard some theoretical constraints,
namely perturbative unitarity of scattering amplitudes in the high energy
limit and
the lack of Landau poles below the Planck scale~\cite{Bause:2021prv}.
In this work, we impose four constraints on the $ B_3 -
L_2 $ model:  limits from direct Higgs searches at hadron colliders, the Higgs
signal strength measurements, agreement with the experimentally measured $ W $
boson mass and perturbativity of the Higgs-quartic couplings at the scale of
the 
effective theory. These correspond to the four coloured regions in
figure~\ref{fig:mixing_exclusion}. 

\subsubsection{Higgs signal strength}
The rotation into the mass eigenbasis of the scalar fields in eq.~\ref{eq:rotation} modifies all SM Higgs couplings by a factor of $ \cos \phi $. The Higgs signal strength, defined as the production cross-section times branching ratio (BR) normalised to the SM prediction,  $ \mu := ( \sigma\times  \text{BR} )_{\text{obs}} / ( \sigma \times \text{BR} )_{\text{SM}}$ for a given Higgs production and decay mode, is then predicted to be $ \mu = \cos^2 \phi $ irrespective of the mode. The ATLAS and CMS Run 2 combination results for the global signal strength read $\mu_{\text{ATLAS}} > 0.92 ,\,  \mu_{\text{CMS}} > 0.90$ at 95 \% CL \cite{Workman:2022ynf}. The more stringent of the two, $ \mu_{\text{ATLAS}} $, yields for the Higgs-flavon mixing angle: $  \left| \sin \phi \right| < 0.28$. This limit on $ \sin \phi $ is independent of the flavon mass.

\subsubsection{Direct searches}
We utilise the public code \textsc{HiggsBounds 5.3.2beta}
\cite{Bechtle:2012lvg, Bechtle:2013wla, Bechtle:2008jh, Bechtle:2011sb,
  Bechtle:2020pkv, Bechtle:2015pma} to obtain 95\% CL direct search limits on
an extra scalar field from the LHC and Tevatron. This bound is stronger
at lower flavon masses and starts to wane for $ m_{\vartheta}
\gtrsim$ 750 GeV; the constraint is visible on the upper left-hand side of
figure~\ref{fig:mixing_exclusion}. 

\subsubsection{Perturbativity of quartic couplings}
For the theory to remain perturbative, we will enforce the conditions $\left|
\lambda_H, \lambda_{\theta}, \lambda_{\theta H} \right| < 4 \pi $ on the
quartic couplings. For most values of the ratio $ x = g_{Z^\prime} \text{~TeV} /
M_{Z^\prime} $ in its allowed region $ x  \in [0.04,0.67]$ the quartic Higgs
coupling $\lambda_{H}$, which is independent of $ x $, places a more stringent bound on $ \sin \phi $ than
either $\lambda_{\theta H}\propto x$ or $ \lambda_\theta \propto x^2$. The
  constraint arising from perturbativity of $ \lambda_H $ corresponds 
to the orange region 
in figure~\ref{fig:mixing_exclusion}. At the end of the allowed interval,
where $ x \sim 0.6 $, the bound from $ \lambda_H $ is superseded by $
\lambda_{\theta H} $.
However, for $m_{\vartheta} \lesssim$ 5~TeV, neither of these limits is competitive against the bound coming from $ M_{W} $ measurements. For flavon masses
much more massive than this, the perturbativity of the couplings becomes the
tightest 
constraint on the mixing angle. This can be phrased in another way: for a
given mixing angle, there is an upper bound on the flavon mass coming from
perturbativity of the three quartic couplings. The strictest bound may depend on $ x $ and the flavon charge $ q_{\theta} $, but $ \lambda_H $ will always provide an upper limit independent of $ x $ and $ q_{\theta} $.

\subsubsection{$W$ boson mass}
We now investigate the prediction of the $ W $ boson mass in the $ B_3 - L_2 $
model. In the real singlet extension of the SM, for an extra scalar field
more massive than $\sim$ 300 GeV, agreement between the experimentally
measured W 
boson mass $ M_{W} $ and the model prediction at the one-loop level places a bound more austere than that arising from the oblique $ S $, $T $ and $ U $ parameters \cite{Lopez-Val:2014jva}. The $ Z^\prime $ boson cannot influence the oblique parameters in the $ B_3 - L_2 $ model, but it does affect $ M_W $ via $ Z^\prime $-induced vertex corrections, as we will see. We thus posit that $ M_W $ will provide the stricter of the two limits in the $ B_3 - L_2 $ model, too, and confirm the assertion by calculating the $ W $ boson mass in the model. 

\begin{figure}[htpb]
	\centering
	\includegraphics[width=0.9\textwidth]{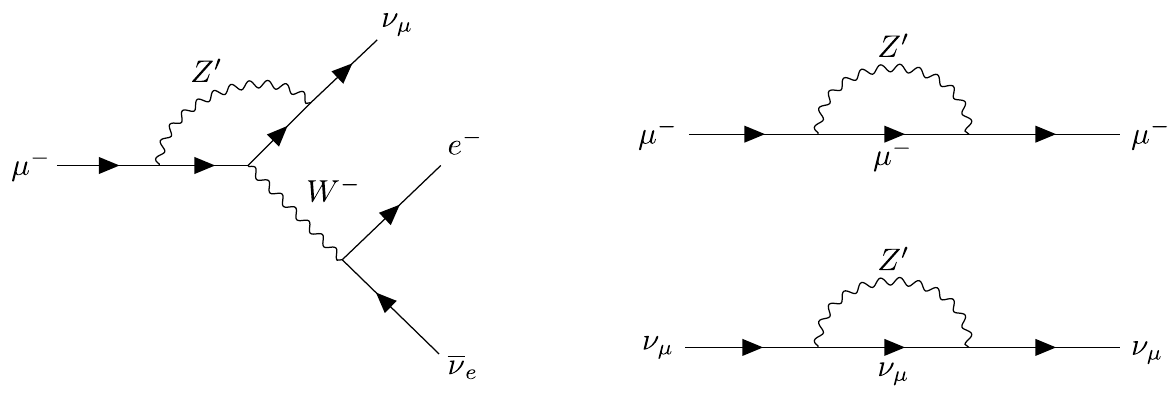}
	\caption{The $ Z^\prime $-induced vertex and self-energy corrections
          contributing to the $ \Delta r $ parameter. 
          There are no contributions
            from similar diagrams but with the 
          flavon Goldstone replacing the $ Z^\prime $ in the loop because
          the $ \ubl $ symmetry is vectorial and does not come with a Yukawa
          sector.}
	\label{fig:zprime_oneloop}
\end{figure}

\begin{figure}
     \centering
     \begin{subfigure}[htpb]{0.3\textwidth}
         \centering
         \includegraphics[width=\textwidth]{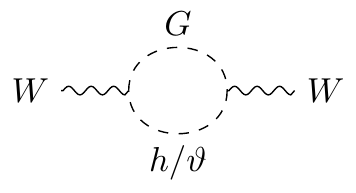}
     \end{subfigure}
     \hfill
     \begin{subfigure}[htpb]{0.3\textwidth}
         \centering
         \includegraphics[width=\textwidth]{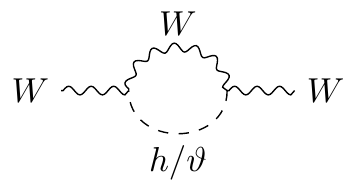}
     \end{subfigure}
     \hfill
     \begin{subfigure}[htpb]{0.3\textwidth}
         \centering
         \includegraphics[width=\textwidth]{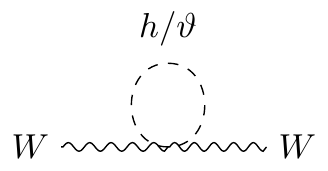}
     \end{subfigure}
        \caption{Feynman diagrams corresponding to the three $W$ boson
          self-energy contributions involving the Higgs boson. We work in the
          Feynman gauge ($\xi = 1$) and $G$ stands for the charged SM Goldstone boson. Each SM $h$-vertex is suppressed by a factor
          of $\cos \phi$ after the Higgs field mixes with the flavon
          field, but this is complemented by a set of identical diagrams
          with the physical flavon field running in the loop. Though not drawn here, the $Z$-boson self-energies are modified in a similar manner.}
        \label{fig:w_self_e}
\end{figure}

Predicting the value of $M_W$ is based on matching the 4-Fermi theory muon lifetime with the 1-loop calculation using the full Lagrangian of the theory (see \cite{Sirlin:1980nh, Hollik:1988ii, Denner:1991kt,Wells:2005vk} for more detailed accounts). The matching yields an expression connecting the experimentally measured Fermi coupling constant $G_F$ to the parameters of the $B_3-L_2$ model:   
\begin{equation}
	\frac{G_F}{\sqrt{2}} = \frac{e^2}{8 M_W^2 \sin^2 \theta_W} \left(1 + \Delta r \right),
\end{equation}
where $\Delta r$ contains all of the loop corrections to the decay process in the full theory. Taking $M_Z$, $G_F$ and $\alpha$ as experimental inputs and working with the on-shell definition of the weak mixing angle where $\sin^2 \theta_W= 1 - M_W^2/M_Z^2$ to all orders in perturbation theory, we can rearrange the above equation to obtain a prediction for the $W$-boson mass:
\begin{equation}
	M_W^2 = 
	\frac{1}{2} M_Z^2 
	\left[ 1+ \sqrt{1 - \frac{4 \pi \alpha}{\sqrt{2} G_F M_Z^2} \left[1 + \Delta r (M_W^2) \right]  } \, \right].
\end{equation}
Given a small perturbation $ \delta (\Delta r) $, resulting from BSM physics, the $ W $ boson mass changes by
\begin{equation}
	\Delta M_W \simeq - \frac{1}{2} M_W \frac{\sin^2 \theta_W}{\sin^2 \theta_W - \cos^2 \theta_W} \delta (\Delta r). \label{eq:deltamw}
\end{equation}
At the one-loop level in the $\ubl$ model, there are two kinds of BSM
contributions to $\Delta r$:\footnote{We neglect loops containing $ \ell \ell h
  $- and $ \ell \ell \vartheta $-vertices, for $ \ell = e,\mu $, as their contributions are of order $  m_{\ell}^2 / v_H^2~\ll~1$.} those arising from
$Z^\prime$-vertices (figure~\ref{fig:zprime_oneloop}) and those arising from
Higgs-flavon mixing (figure~\ref{fig:w_self_e}), which were evaluated in
ref.~\cite{Lopez-Val:2014jva} and will always act to make the $ W $ boson
lighter. Each of the two sets of diagrams depends on different parameters: the size of the $Z^\prime$ vertex contributions is a function of $\{M_{Z^\prime}, g_{Z^\prime}\}$ whereas Higgs-flavon mixing hinges on $\{ \sin \phi, m_{\vartheta} \}$.

We employ the \textsc{FeynArts, FormCalc} and \textsc{LoopTools} packages
(versions 3.11, 9.9 and 2.16, respectively) \cite{Hahn:2000kx, Hahn:1998yk} to
aid with the calculation and to evaluate the results numerically. Ignoring
terms of order $ m_{\mu}^2 / M_W^2$ and $ m_{\mu}^2 / M_{Z^\prime}^2 $, the
Lorentz structure of the $ \ubl $ amplitude is identical to that of the
4-Fermi theory and we can match the the two theories at the amplitude
level. In doing so, we find that $ \delta (\Delta r) $ is dominated by the
oblique corrections to the vector boson propagators, overwhelming the $
Z^\prime $-induced effects by several orders of magnitude. The reason is that
all of the $ Z^\prime $-effects are of order $ g_{Z^\prime}^2 m_{\mu}^2  / M_{Z^\prime}^2 $ and thus become negligible for a~TeV scale $ Z^\prime $. Thus, the resulting constraint on $ \sin \phi $ is essentially independent of $ M_{Z^\prime} $ and $ g_{Z^\prime} $.

We proceed by comparing the theoretical prediction with the experimentally
determined $  W$ boson mass and insist on agreement at the 2$ \sigma $
level. The empirical and SM-predicted values of $M_W$ are obtained from the Particle Data Group
\cite{Workman:2022ynf}. For 
the experimentally measured $ M_W $, we use the current world average
(prior to the 2022 CDF measurement\footnote{If the 2022 CDF measurement is considered, all non-zero mixing angles are ruled out at the $ 2  \sigma $ level. The $\ubl$ model can only make the $ W $ boson lighter than predicted by the SM, thus \emph{increasing} the tension between experiment and theory.}), $ M_W^{\text{exp}} = 80.377 \pm 0.012$
GeV, whereas the SM prediction stands at $  M_W^{\text{th} }= 80.356 \pm
0.006$ GeV. Combining the two errors in quadrature, we use
eq.~\ref{eq:deltamw} to obtain a $ \ubl $ model prediction for the mass of the
$ W $ boson and require that the predicted and measured values disagree by
less than $ 2\sigma $. This constraint corresponds to the yellow region in
figure~\ref{fig:mixing_exclusion} and is the most stringent for
$0.5<m_\vartheta/\text{~TeV}<5$.

\subsection{Flavon decay channels}
\begin{figure}[htpb]
	\centering
	\includegraphics[width=0.8\textwidth]{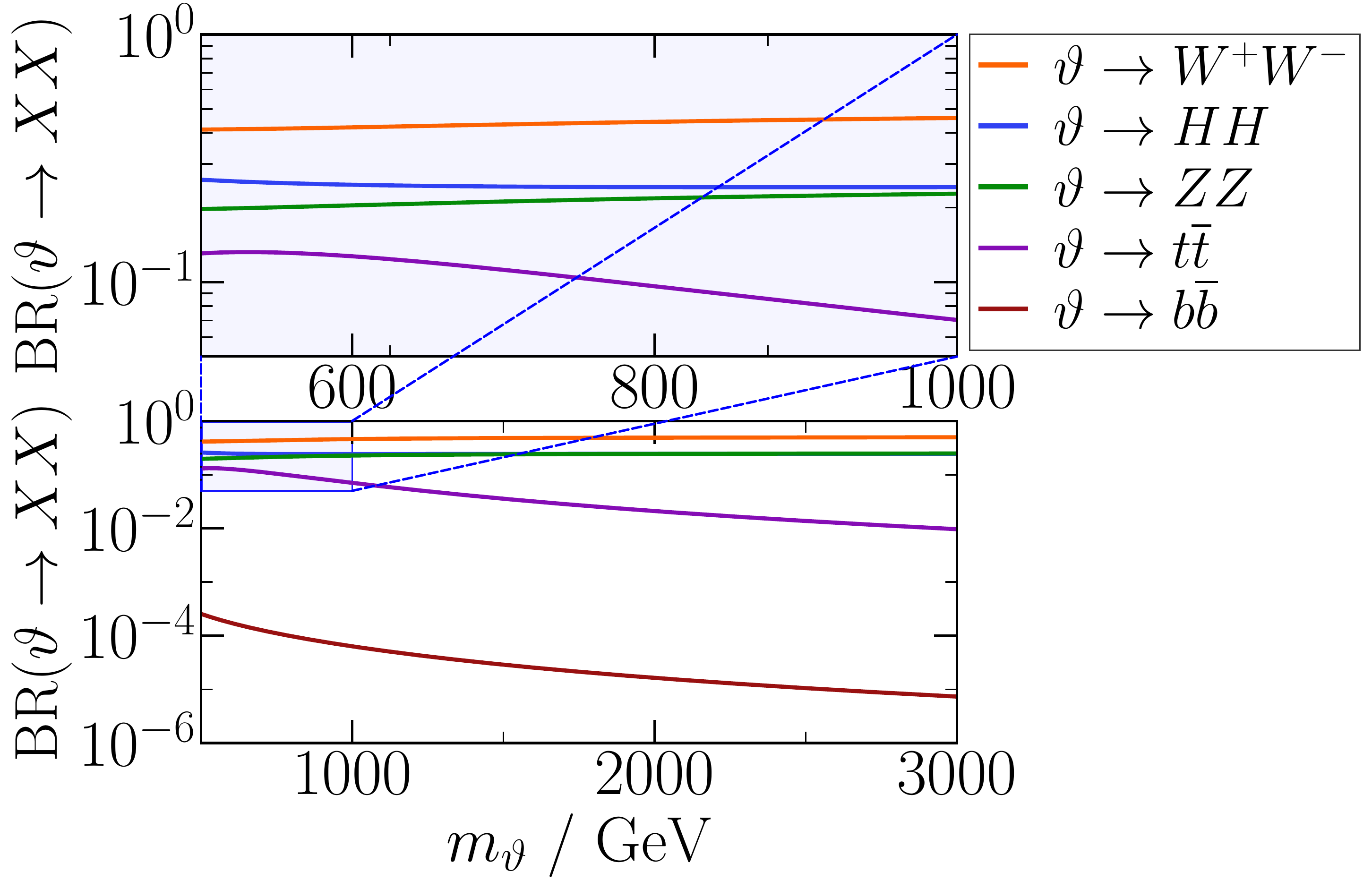}
	\caption{Tree-level flavon BRs for the case $ m_{\vartheta} < 2
	M_{Z^\prime} $ so that the flavon is unable to decay into a pair of on-shell
	$ Z^\prime$ bosons. The three leading final states, $ WW $, $ ZZ $ and $ hh$, come in the approximate ratio 2:1:1. The upper panel is an enlargened version
	of the shaded region of the lower panel.  \label{fig:flavon_BR}}
\end{figure}

As the flavon couples to SM fields only through mixing with the Higgs
field, its tree-level decay channels resemble those of the SM Higgs whenever $ \sin \phi \neq 0 $. A key difference, however, is that the flavon
is assumed to be considerably heavier than the Higgs, which allows decays into
on-shell $  W^-W^+,ZZ$ and $t\overline{t} $ final states. Assuming a~TeV scale
flavon and $Z^\prime$, and that $ m_{\vartheta} < 2 M_{Z^\prime} $ (which
covers most of the parameter space studied in this work), there are three
channels that dominate the flavon decay rate. The leading channel is $\vartheta \rightarrow W W$ with tree-level partial width
\begin{equation}
	\Gamma_{\vartheta \rightarrow W W} = \frac{m_{\vartheta}^{3} \sin^2{\phi} }{16 \pi v_{H}^{2}} + \mathcal{O}\left(\frac{M_W^2}{m_{\vartheta}^2} \right).
\end{equation}
This is followed by $ \vartheta \rightarrow h h $ (obtained with \textsc{FeynRules}): 
\begin{equation}
	 \Gamma_{\vartheta \rightarrow h h} = \frac{m_{\vartheta}^{3} \cos^{4} \phi \sin^{2} \phi}{32 \pi v_{H}^{2}} + \mathcal{O}\left(\frac{M_W^2}{m_{\vartheta}^2} \right) +  \mathcal{O} \left( \frac{v_H}{v_{\theta}} \right)
\end{equation}
and $ \vartheta \rightarrow ZZ $:
\begin{equation}
	\Gamma_{\vartheta \rightarrow Z Z} = \frac{m_{\vartheta}^{3} \sin^2{\phi} }{32 \pi v_{H}^{2}} + \mathcal{O}\left(\frac{M_Z^2}{m_{\vartheta}^2} \right).
\end{equation}
These expressions lead to the relation $\Gamma_{\vartheta \rightarrow W W} /
\Gamma_{\vartheta \rightarrow h h} \approx 2 \approx \Gamma_{\vartheta
  \rightarrow W W} / \Gamma_{\vartheta \rightarrow Z Z} $, which is clearly
demonstrated in figure~\ref{fig:flavon_BR}.
The leading fermionic final state is a $t \overline{t}$ pair because the top
  quark Yukawa coupling is the largest Yukawa coupling in the SM\@.
 We have evaluated the BRs numerically using \textsc{MadWidth} \cite{Alwall:2014bza}. 

\begin{figure}[htpb]
	\centering
	\includegraphics[width=0.7\textwidth]{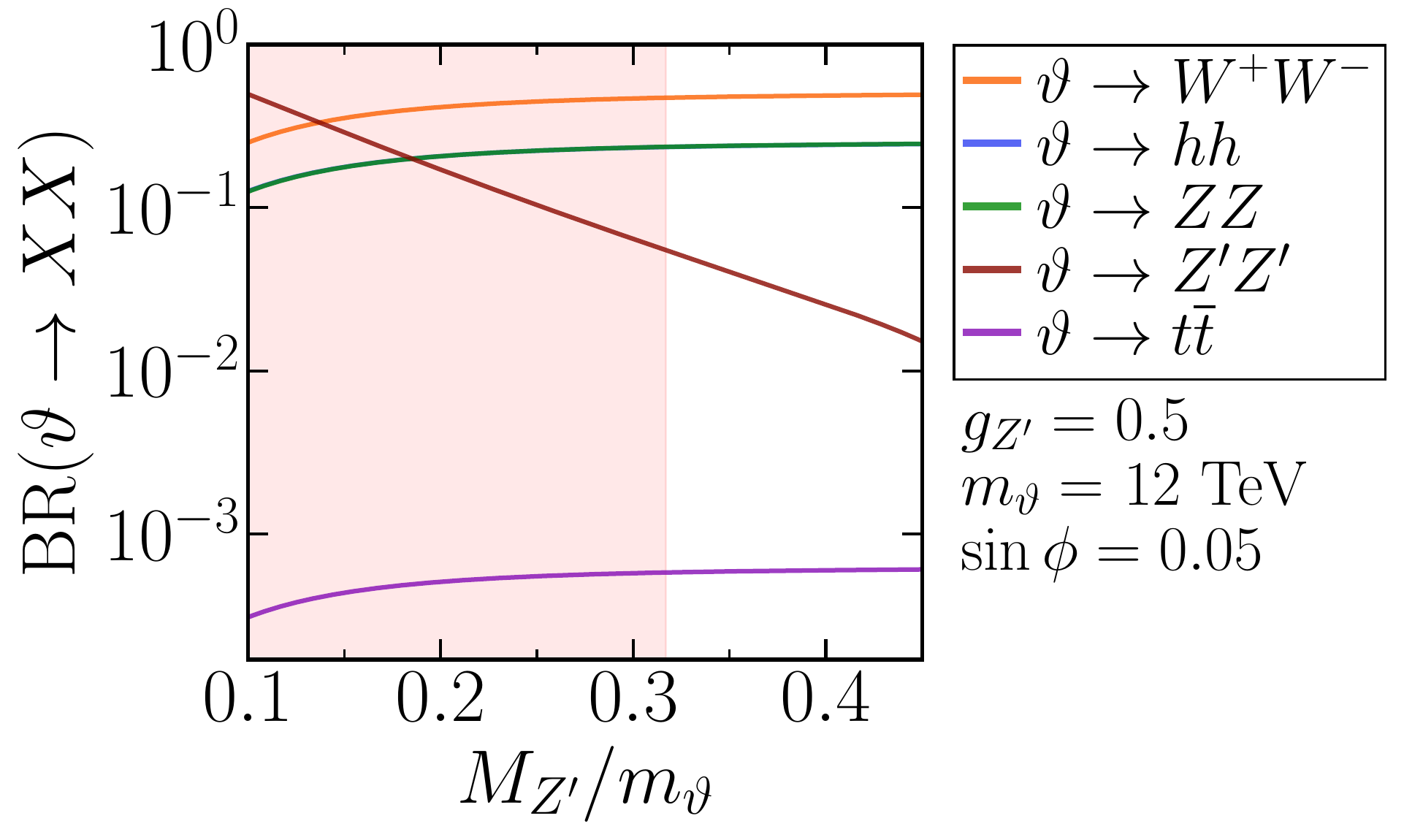}
	\caption{Flavon BRs as a function of the mass ratio $ M_{Z^\prime} /
          m_{\vartheta} $ when the flavon is much heavier than the $ Z^\prime
          $ and the $ \ubl $ gauge coupling is order one. As the mass ratio is
  lowered, we see that the $ Z^\prime Z^\prime $ final state becomes increasingly important. The region excluded at the 95\% CL by LHC data, for this particular choice of parameters, is shown in pink. The green and blue lines are overlap to such a degree that they are indistinguishable by eye.}
	\label{fig:high_flav_mass}
\end{figure}

We may also study the case with $ m_{\vartheta} > 2 M_{Z^\prime} $, although,
owing to the constraint $ M_{Z^\prime} \gtrsim$ 2~TeV, this necessarily takes
us to multi-TeV flavon masses. If one allows for a flavon mass of order
10~TeV, the decay $ \vartheta \rightarrow Z^\prime Z^\prime $ can become one
of the leading channels. This is shown in figure~\ref{fig:high_flav_mass}
where the BR into a pair of $ Z^\prime $s keeps increasing rapidly as more
phase space is made available by lowering $ M_{Z^\prime}/m_{\vartheta} $. To
obtain the figure, we have arbitrarily picked $  m_{\vartheta} = 12$~TeV, $\sin
\phi = 0.05$ and $ g_{Z^\prime} = 0.5 $. The pink region is excluded at the
95\% CL by LHC data, as shown in figure 10 of ref. \cite{Azatov:2022itm}. This
constraint, which is a function of $ g_{Z^\prime} / M_{Z^\prime} $, is
included for completeness only --- the primary purpose of the figure is to show
how the mass ratio influences the BRs.

\section{Flavonstrahlung at Colliders \label{sec:prod}}
\begin{figure}[htbp]
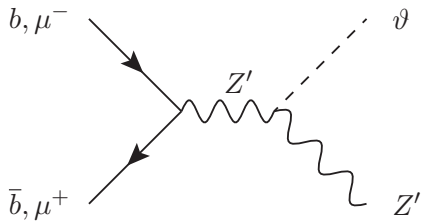

	\begin{center}
		\SetScale{1.4}
	  \begin{axopicture}(300,75)(100,0)
	    \Line[arrow](220,50)(245,25)
	    \Line[arrow](245,25)(220,0)
	    \DashLine(295,50)(270,25){3}
	    \Photon(270,25)(295,0){3}{3}
	    \Photon(245,25)(270,25){3}{3}
	    \Text(260.5,33)[c]{${Z^\prime}$}
	    \Text(215,0)[r]{$\overline{b}, \mu^+$}
	    \Text(215,50)[r]{$b, \mu^-$}
	    \Text(302,50)[l]{$\vartheta$}
	    \Text(302,0)[l]{$Z^\prime$}
	    \end{axopicture}
	\end{center}
	\caption{Flavonstrahlung at a hadron collider or a muon collider.\label{fig:flavonstrahlung}}
\end{figure}

We now proceed to study the tree-level flavonstrahlung cross-section at hadron and muon colliders of varying centre-of-mass energies (the dominant Feynman diagram is shown again in figure~\ref{fig:flavonstrahlung}). 
Flavonstrahlung would likely not be the first detectable direct BSM signal in the $ \ubl $ model, as it is more probable that exclusive $ Z^\prime $ production would be observed at lower energies and luminosities. 
As to whether flavonstrahlung would be the first sign of the flavon particle depends primarily on the value of the Higgs-flavon mixing angle. For sizeable mixing angles, we may discover the flavon through the conventional SM Higgs production processes before reaching the energies and luminosities required for flavonstrahlung.
However, observations of $ Z^\prime $ and flavon resonances alone would not tell us whether the
two particles interact with each other and whether the scalar field is
involved in generating the $ Z^\prime $ mass.
Flavonstrahlung is unique in that it combines  the $ Z^\prime $ and flavon in a single process.
The subsequent decays of the $ Z^\prime $ and $ \vartheta $ via their leading
channels, $ Z^\prime \rightarrow \mu^- \mu^+ $ and $ \vartheta \rightarrow W^-
W^+ $, yield the final state $ W^+ W^- \mu^+ \mu^- $ with a $ WW $ resonance
at the flavon mass $ m_{\vartheta} $ and a di-muon resonance
at $ M_{Z^\prime} $. 
As we have seen, the flavon may also decay to $ ZZ $ or $ HH $ with
sizeable BRs, and we leave it for future work to determine which channel is
best for flavonstrahlung hunting.  

\begin{figure}[htpb]
	\centering
	\includegraphics[width=0.9\textwidth]{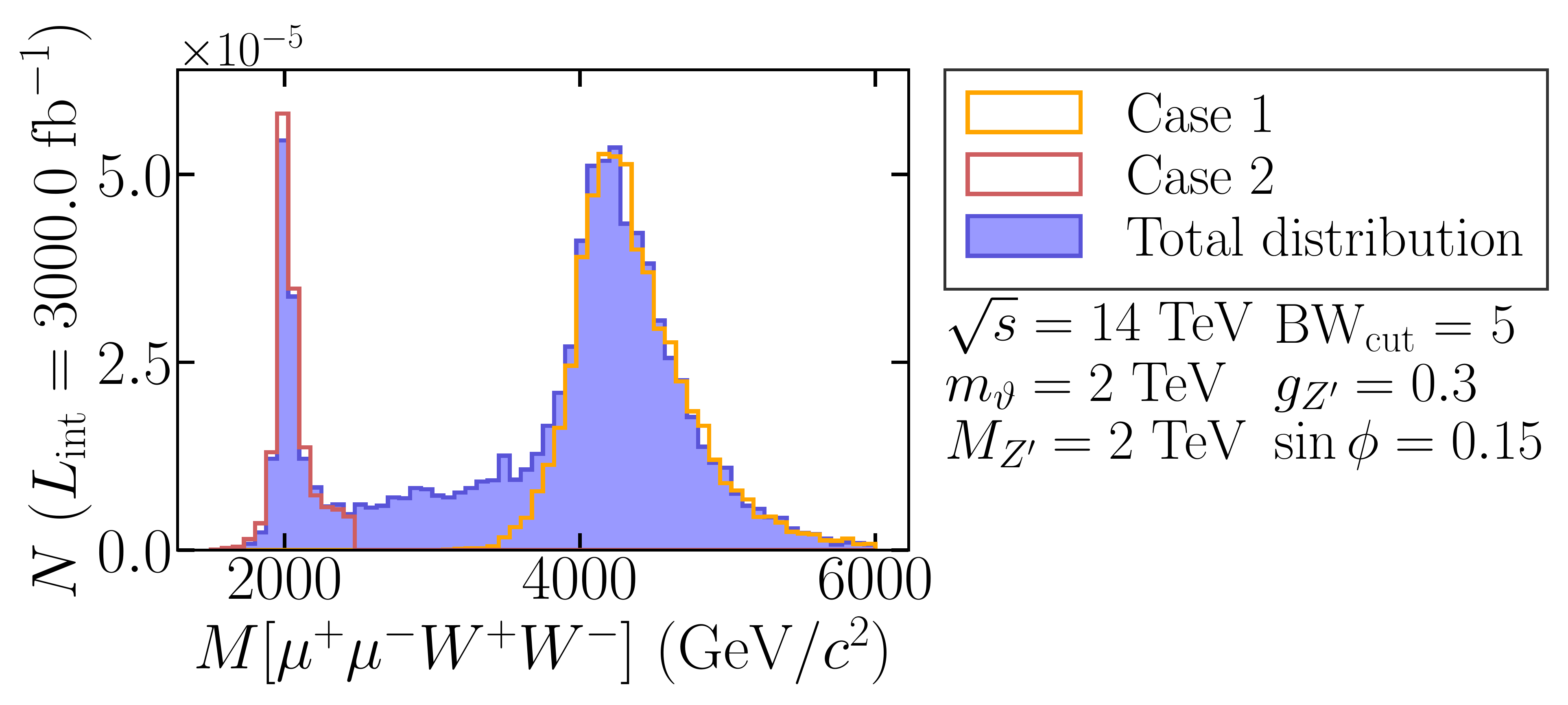}
	\caption{Illustration of the two resonant contributions to the flavonstrahlung cross-section for the leading $ \mu^+ \mu^- W^+ W^- $ final state. The blue region shows the final state invariant mass distribution in the absence of SM backgrounds at the 14 TeV HL-LHC\@. The orange curve delineates the contribution arising from case 1, where the second $ Z^\prime $ in figure~\ref{fig:flavonstrahlung} is on-shell, whereas the red curve corresponds to case 2, where the first $ Z^\prime $ is on-shell.  }
	\label{fig:on_shell}
\end{figure}

It should be noted that there are two resonant contributions to the flavonstrahlung cross-section: in addition to $ p p \rightarrow {Z^\prime}^\ast \rightarrow Z^\prime \vartheta $ (case 1), discussed above, the cross-section also picks up a contribution from $ p p \rightarrow Z^\prime \rightarrow {Z^\prime}^\ast \vartheta $ (case 2), where the intermediate $ Z^\prime $ is on-shell. 
Choosing for concreteness the leading decay channels $ Z^\prime \rightarrow \mu^- \mu^+ $ and $ \vartheta \rightarrow W^- W^+ $, case 2 yields a resonance peak in the invariant mass of the 4-particle final state at $ q^2~=~M_{Z^\prime}^2 $ as opposed to the di-muon resonance in case 1. 
We use the event generator \textsc{\mbox{MadGraph5\_aMC@NLO}} v.3.4.1  \cite{Alwall:2014hca} (abbreviated ``\textsc{MG5\_aMC}'' in the following) to illustrate the two resonances schematically in figure~\ref{fig:on_shell} for the final state $ \mu^+ \mu^- W^+ W^- $ at a hadron collider. 
The orange and red lines corresponds to cases 1 and 2, respectively. 
We have used the condition $ | M^\ast - M | < 5 \Gamma $ as the definition of a propagator being on-shell, with $ M^\ast $ the invariant mass of the four-momentum carried by the propagator and $ M $ and $ \Gamma $ the pole mass and width of the particle. 
This is achieved using the $ \text{BW}_\text{cut} $ parameter in \textsc{MG5\_aMC}.

As to which resonance contributes more to the cross-section depends on the
partonic energy and the masses of the flavon and $ Z^\prime $. Because case 2
entails lower centre-of-mass energies than case 1, it is more prevalent at
lower energy colliders such as the HL-LHC or a 3 TeV muon collider which may
lack high enough partonic energies to put both a TeV scale $ Z^\prime $ and $
\vartheta$ on-shell simultaneously. Case 1, on the other hand, is favoured at
higher energy colliders such as the FCC-hh or a 10 TeV muon collider where
the sum of the $Z^\prime$ and
  $\vartheta$ masses is less than the partonic centre of mass energy for a
  substantial fraction of collisions.


In order to capture contributions from both resonances in our cross-section computations, we shall study the process $ p p \rightarrow Z' \rightarrow \vartheta Z'  \rightarrow \vartheta \mu^+ \mu^- $, where the rightmost $ Z'$ propagator in the Feynman diagram of figure~\ref{fig:flavonstrahlung} splits into a di-muon pair.\footnote{All flavonstrahlung cross-sections reported hereafter are computed for the $ \vartheta \mu^+ \mu^- $ final state, even when not explicitly stated.} 
We require only that the final state flavon and muons are on-shell, thus allowing both case 1 and case 2 from above to contribute to the total amplitude. 
Concentrating on the di-muon final state is well-motivated by its clean experimental signature, as well as the large $ Z' \rightarrow \mu^+ \mu^- $ branching ratio, making it the most promising mode for observing flavonstrahlung.

\begin{figure}[htbp]
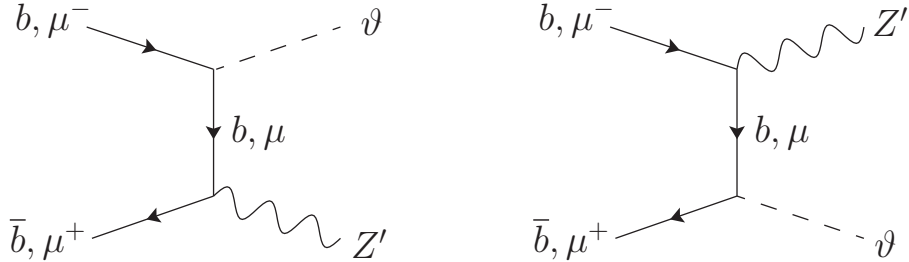

	\begin{center}
		\fontsize{14}{14}
	  \begin{axopicture}(98,86) (175,-95)
	    \SetColor{Black}
	    \Line[arrow,arrowpos=0.5,arrowlength=5,arrowwidth=3,arrowinset=0.2](128,-10)(176,-26)
	    \Line[arrow,arrowpos=0.5,arrowlength=5,arrowwidth=3,arrowinset=0.2](176,-26)(176,-74)
	    \Line[arrow,arrowpos=0.5,arrowlength=5,arrowwidth=3,arrowinset=0.2](176,-74)(130,-90)
	    \Photon(176,-74)(224,-90){5}{3}
	    \Line[dash,dashsize=6.2](177,-26)(224,-10)
	    \Text(193,-50)[c]{$b, \mu$}
	    \Text(130,-7)[r]{$b, \mu^-$}
	    \Text(128,-90)[r]{$\overline{b},\mu^+$}
	    \Text(240,-8)[r]{$\vartheta$}
	    \Text(228,-92)[l]{$Z^\prime$}
	  \end{axopicture}
		\begin{axopicture}(98,86) (75,-95)
	    \SetColor{Black}
	    \Line[arrow,arrowpos=0.5,arrowlength=5,arrowwidth=3,arrowinset=0.2](128,-10)(176,-26)
	    \Line[arrow,arrowpos=0.5,arrowlength=5,arrowwidth=3,arrowinset=0.2](176,-26)(176,-74)
	    \Line[arrow,arrowpos=0.5,arrowlength=5,arrowwidth=3,arrowinset=0.2](176,-74)(130,-90)
	    \Photon(176,-26)(224,-10){5}{3}
	    \Line[dash,dashsize=6.2](176,-74)(224,-90)
	    \Text(193,-50)[c]{$b, \mu$}
	    \Text(130,-7)[r]{$b, \mu^-$}
	    \Text(128,-90)[r]{$\overline{b},\mu^+$}
	    \Text(241,-8)[r]{$Z^\prime$}
	    \Text(228,-92)[l]{$\vartheta$}
	  \end{axopicture}
	\end{center}
	\caption{Flavon production with an associated $ Z^\prime $ via a $ t $-channel fermion exchange. The bottom quark exchange corresponds to hadron colliders, whereas muon exchange is possible at muon colliders. For the regions of parameter space considered in this work, the contribution to the $ \vartheta Z^\prime $ production cross-section from this channel is typically of order $ 0.1\% $ or less. \label{fig:t_channel}}
\end{figure}

In the presence of non-zero Higgs-flavon mixing, a $ Z^\prime \vartheta $ final state may also be produced at tree-level via a $ t $-channel bottom quark or muon exchange. These processes are shown in figure~\ref{fig:t_channel}. The associated matrix elements are suppressed because the $ \vartheta b b $- and $ \vartheta \mu \mu $-vertices come with couplings $ \sin \phi \left( m_b / v_H \right)  $ and  $ \sin \phi \left( m_\mu / v_H \right) $, respectively. Assuming $\sin \phi \sim 0.1  $, the inclusion of the $ t $-channel exchange typically changes the $ \vartheta Z^\prime $ production cross-sections by approximately $0.1\%$ and never by more than around $ 3\%$. We thus neglect contributions arising from the $ t $-channel fermion exchange in this work. 

\subsection{Flavonstrahlung at Hadron Colliders}

We import the UFO model file into \textsc{MG5\_aMC} and use it to calculate
leading-order flavonstrahlung cross-sections for proton-proton ($ pp $)
collisions. The largest partonic contribution to $ Z^\prime $ production comes
from the $ b \overline{b} $ initial state. We thus use the five-flavour parton
distribution function (PDF) NNPDF2.3LO where the $ b $ quark is absorbed into
the proton and jet definitions and treated as massless. There are also negligible contributions to the cross-sections
from $ s \overline{b} $, $ b \overline{s} $ and $ s \overline{s} $ initial
states, which are nevertheless included in our numerical estimates. 

We apply the default \textsc{MG5\_aMC} cuts on the phase space of the
  di-muon pair throughout the computations. Placing cuts according to the
  specifications of each detector studied in this work would have a
  negligible impact on our results, and the  designs of future detectors
    are not yet fixed anyway. We thus require that the final-state muon isolation satisfy $ \Delta R > 0.4 $ and that transverse momenta and pseudo-rapidities of the muons fulfil the conditions $ p_T > 10 $~GeV and $ \left| \eta \right| <2.5 $. No cuts are applied on the final state flavon.

\begin{figure}[t]
\begin{center}
\unitlength=15cm
\begin{picture}(1,0.45)(0,0)
    \put(-0.03,0.05){\includegraphics[width=0.48\textwidth]{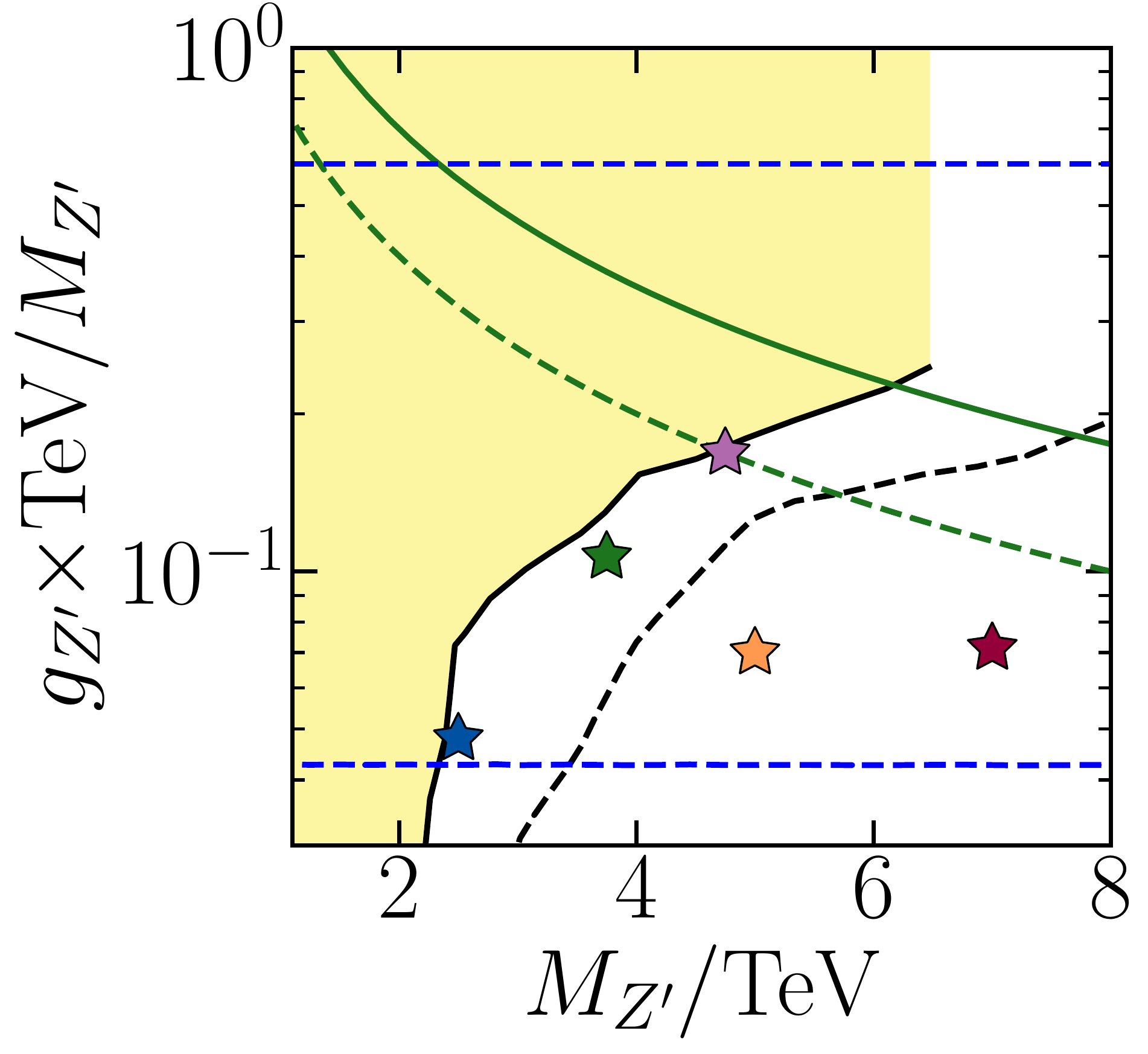}}
    \put(0.45,-0.05){\includegraphics[width=0.55\textwidth]{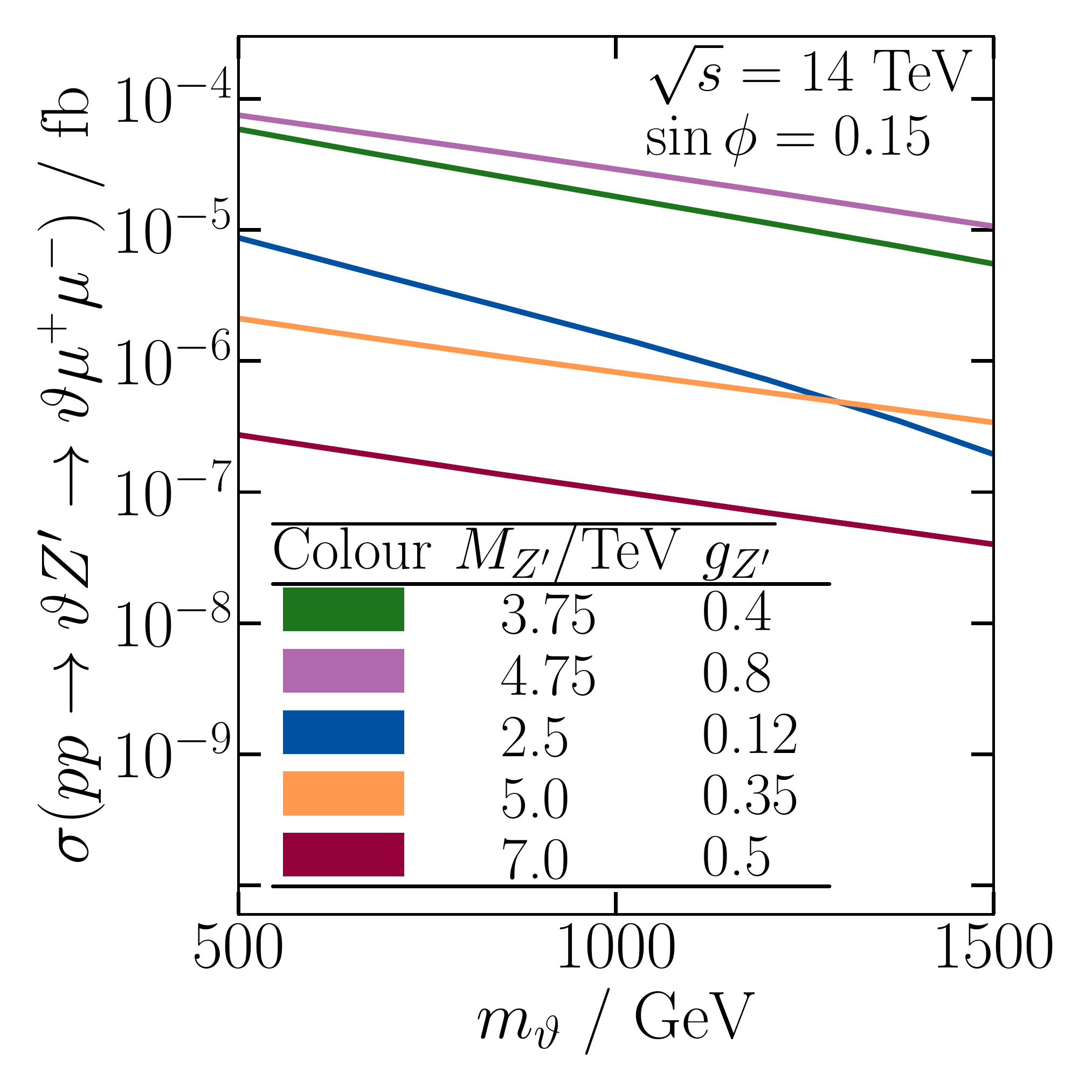}}
\end{picture}
\end{center}
\caption{The left-hand panel, based on figure 9 of ref.~\cite{Azatov:2022itm}, shows
    the $ g_{Z^\prime} - M_{Z^\prime} $ plane of the parameter space. Everything above the
    solid black line is excluded at the 95\% CL by the LHC whereas the dashed black line
    indicates the projected 95\% CL sensitivity of the HL-LHC\@. The dashed and solid green lines indicate the $ \Gamma / M_{Z'} = 1 / 3 $ and $  \Gamma / M_{Z'} = 1 $ bounds, above which perturbative computations become inaccurate. The region between the blue
    dashed lines is the region allowed by the fits discussed in
    section~\ref{sec:fit_to_NCBA}. Were another fit including the new
      $R_K$ and $R_{K^\ast}$ measurements~\cite{LHCb:2022qnv, LHCb:2022zom} to be performed, the
      position of the lower blue line would be revised downward. Coloured stars have been superposed on the figure, with each star labelling a benchmark point in the parameter plane. The right-hand panel shows tree-level flavonstrahlung cross-sections for 14~TeV $ pp $ collisions with the flavon charge $ q_{\theta}$ set to unity. Each coloured line corresponds to a parameter space point labelled by a star of the same colour. \label{fig:14TeV_combined}}
\end{figure}

To study the process, we select currently allowed combinations of $
\{M_{Z^\prime}, g_{Z^\prime}  \} $, and compute the flavonstrahlung
cross-section as a function of the flavon mass. These combinations are
illustrated in the left-hand panel of figure~\ref{fig:14TeV_combined} which is
adapted from figure~10 of ref. \cite{Azatov:2022itm}. The region above the
solid black line has been excluded at the 95\% CL by the CMS high-mass Drell-Yan searches. The dashed blue lines mark the condition $0.04
< x < 0.67  $, whereas the green lines delineate perturbativity
conditions. We have added five coloured stars representing example
parameter combinations to the plot for which we compute benchmark
cross-sections. A representative 
value of the Higgs-flavon mixing angle, $ \sin \phi = 0.15 $, is chosen in all
simulations, keeping in mind that the flavonstrahlung cross-section is
proportional to $ \cos^2 \phi $. We also keep the flavon charge $ q_{\theta} $
at unity for now. 

We first consider the cross-section $ \sigma\left( p p \rightarrow
\vartheta Z^\prime  \rightarrow \vartheta \mu^+ \mu^-   \right)  $ at a centre-of-mass energy $ \sqrt{s} =$ 14~TeV, representing the HL-LHC\@. The cross-sections for the five example points are
shown in the right-hand panel of figure~\ref{fig:14TeV_combined}, where the
colours of the lines correspond to the colours of the stars in the left-hand
panel. The exact choices of $ \{M_{Z^\prime}, g_{Z^\prime}\} $ are listed in the
legend. Assuming HL-LHC integrated luminosity of 3000 $\text{fb}^{-1}$, the plot
suggests we expect to produce less than $ \mathcal{O}(1) $ flavonstrahlung events. Thus, the flavonstrahlung cross-sections are too small for discovery at the HL-LHC.

\begin{figure}[htpb]
	\centering
	\includegraphics[width=0.75\textwidth]{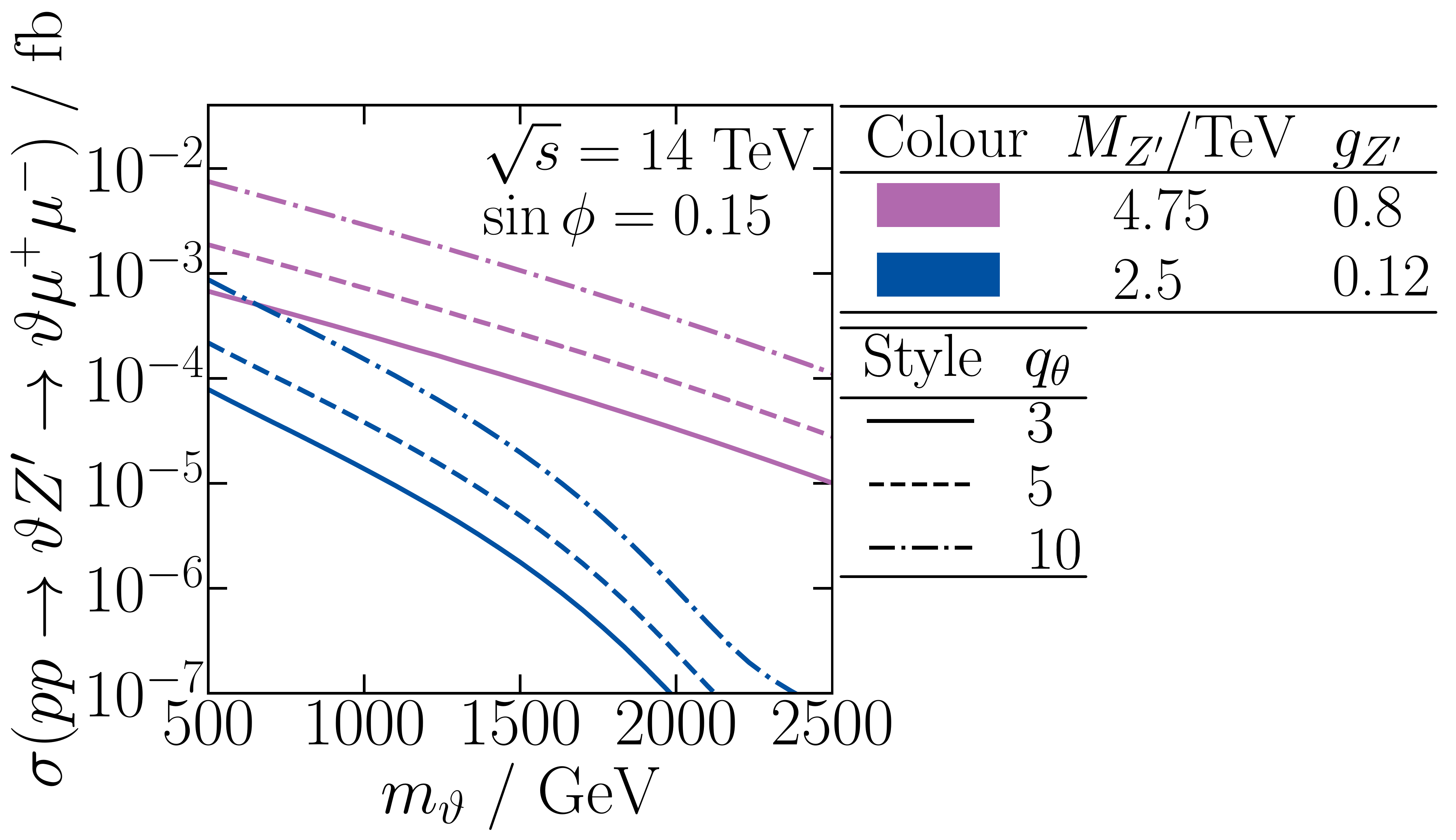}
	\caption{Flavonstrahlung cross-sections for 14~TeV $ pp $ collisions
          but with a variable flavon charge. The charge $ q_{\theta}$ takes on
          values 3, 5 and 10, represented by solid, dashed and dash-dotted
          lines, respectively. The line colours represent different points in
          the $ M_{Z^\prime} $ -- $ g_{Z^\prime} $ plane and are congruent
          with the colours of the stars in
          figure~\ref{fig:14TeV_combined}.} 
	\label{fig:qth_comparison}
\end{figure}

Figure~\ref{fig:qth_comparison} shows how the picture changes if the flavon charge $ q_{\theta}  $, which can be any rational number, is varied.
We have selected the two parameter combinations from
figure~\ref{fig:14TeV_combined} which yield the largest and third largest
cross-sections and let $ q_{\theta} $ take values 3, 5 and 10. 
 We find that for flavon charges $ q_{\theta} \gtrsim 5 $ and for $ Z^\prime $ masses and couplings near the current exclusion limits, the HL-LHC may be able to discover flavonstrahlung up to around 1~TeV flavon masses, but a detailed study would be necessary to confirm this.
Either way, the above relies on a finely tuned selection of input parameters
and does not change the overall conclusion the HL-LHC lacks sufficient
partonic energies and luminosity to look for flavonstrahlung in more than a
corner of the currently available parameter space. For the rest of the
  present paper, we return to flavon charges of unity.

\begin{figure}[htpb]
	\centering
	\includegraphics[width=0.7\textwidth]{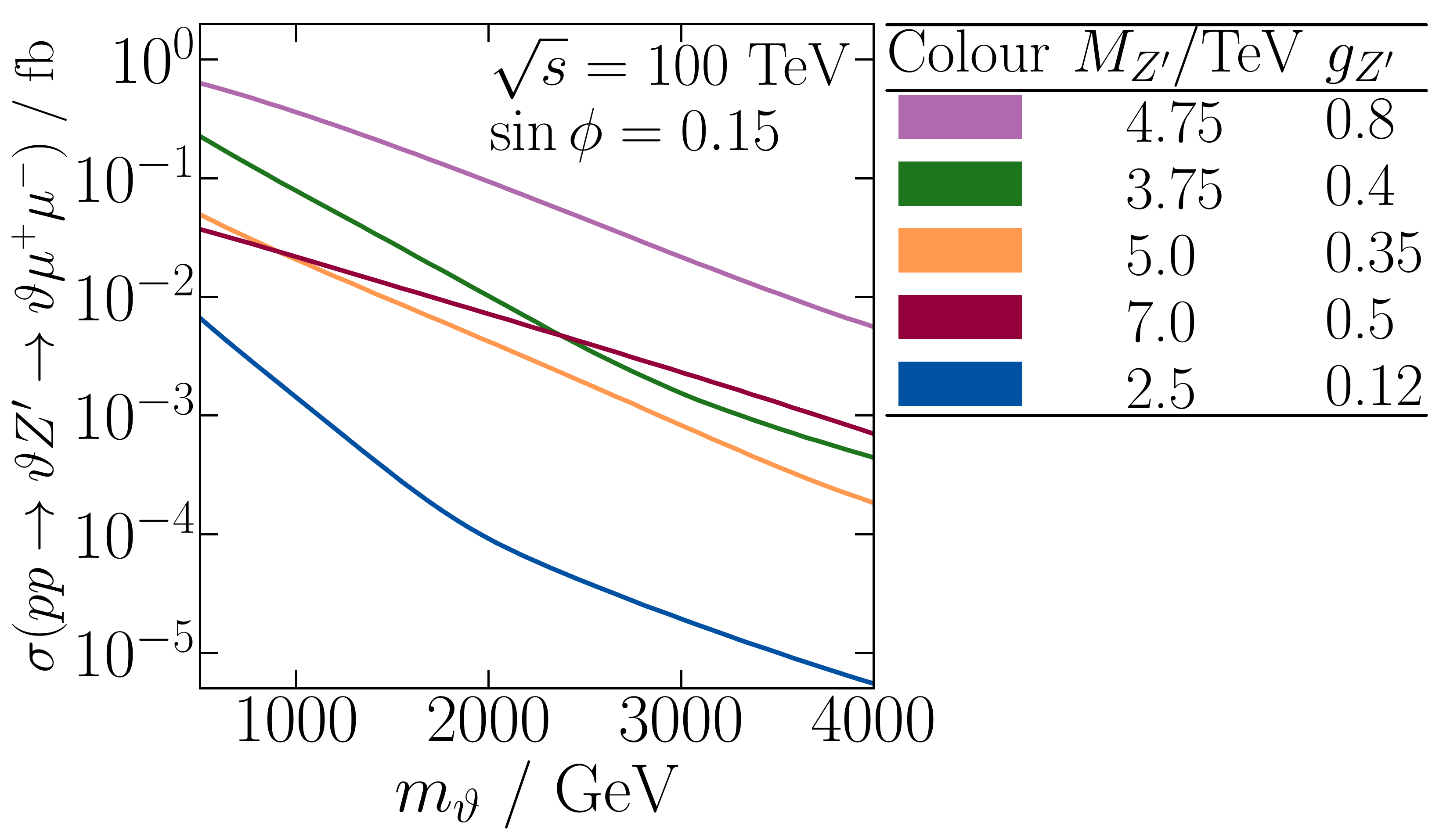}
	\caption{Tree-level flavonstrahlung cross-sections for 100~TeV $ pp $
          collisions for $q_\theta=1$. Each coloured line corresponds to
          a parameter space point labelled by a star of the same colour in
          figure~\ref{fig:14TeV_combined}. \label{fig:100TeV_combined}} 
\end{figure}

We now examine whether a 100~TeV hadron collider, such as the FCC-hh with an
integrated luminosity of 20--30 $\text{ab}^{-1}$, would be capable of
discovering flavonstrahlung. The simulated cross-sections for the the
five parameter space points indicated by the
  coloured stars in figure~\ref{fig:14TeV_combined} are shown in figure~\ref{fig:100TeV_combined}. The resulting cross-sections are greatly enhanced compared to HL-LHC and the larger partonic energies come with the benefit that the cross-sections are not as sensitive to the flavon and $ Z^\prime $ masses. 

To gain some insight into the reach of the collider, we discard as undiscoverable those parameter space points where less than 10 flavonstrahlung events are expected to be produced. Parameter space points passing this very rough criterion are not automatically within the reach of the collider, and we leave for a future work the detailed study of the SM backgrounds and detector effects which would allow for a more precise estimate of the collider sensitivity. Applying this crude method, we find that for all but the smallest allowed values of the gauge coupling, $ g_{Z^\prime}\gtrsim 0.4$, the collider can explore the parameter space up to $ \sim$ 5~TeV flavon and $ Z^\prime $ masses. For $ g_{Z^\prime} \lesssim  0.4$, the mass reach will likely be more limited. The blue line in figure~\ref{fig:100TeV_combined}, corresponding to the smallest allowed coupling for a 2.5~TeV $ Z^\prime $ boson, demonstrates that even in this case we may have sensitivity up to a flavon mass of around 2~TeV.

\subsection{Flavonstrahlung at Muon Colliders}
\begin{figure}[htpb]
	\centering
	\includegraphics[width=0.8\textwidth]{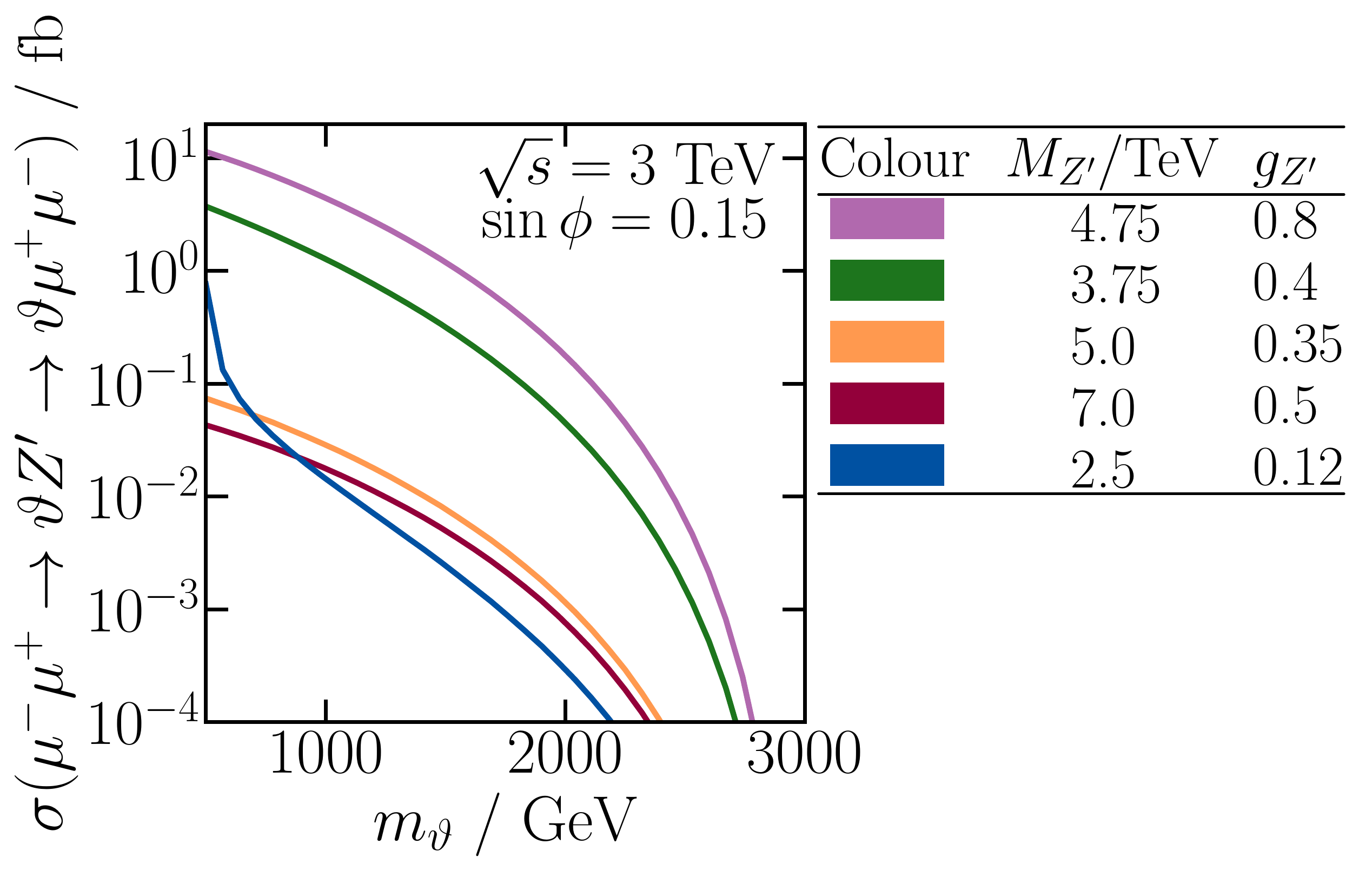}
	\caption{Tree-level flavonstrahlung cross-sections for 3~TeV $ \mu^+ \mu^- $ collisions for $ q_{\theta} = 1 $. Each coloured line corresponds to a parameter point labelled by a star of the same colour in figure~\ref{fig:14TeV_combined}. The cross-sections do not include ISR effects.}
	\label{fig:muon_collider_3}
\end{figure}

\begin{figure}[htpb]
	\centering
	\includegraphics[width=0.8\textwidth]{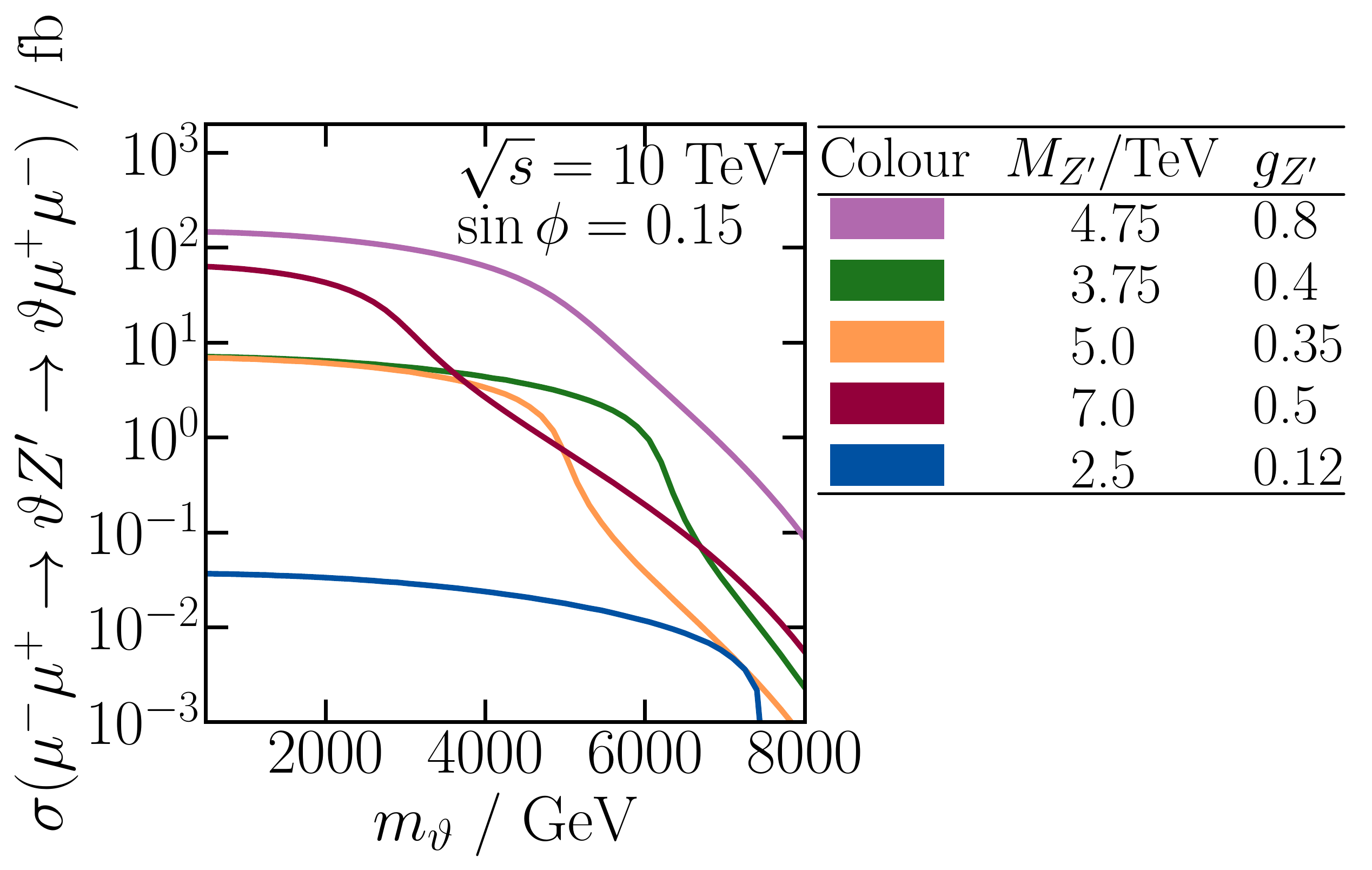}
	\caption{Tree-level flavonstrahlung cross-sections for 10~TeV $ \mu^+ \mu^- $ collisions for $ q_{\theta} = 1 $. Each coloured line corresponds to a parameter point labelled by a star of the same colour in figure~\ref{fig:14TeV_combined}. The cross-sections do not include ISR effects.}
	\label{fig:muon_collider_10}
\end{figure}

We may also simulate flavonstrahlung at 3~TeV 
and 10~TeV $ \mu^+ \mu^- $
colliders assumed to reach integrated luminosities of 1~$\text{ab}^{-1}$ and
10 $\text{ab}^{-1}$, respectively.
Despite the centre-of-mass energies being lower than that of the FCC-hh, the fact that
nearly all of the beam energy is typically carried by the colliding muon pair may allow
these colliders to have high sensitivity to flavonstrahlung. We once again
focus on the five example points from figure~\ref{fig:14TeV_combined} and
use \textsc{MG5\_aMC} to perform the simulations. The results for the 3 TeV and 10 TeV colliders are shown in
figures \ref{fig:muon_collider_3} and \ref{fig:muon_collider_10}, respectively.  

The cross-sections do not include initial state radiation (ISR) effects, as this feature is not yet implemented in \textsc{MG5\_aMC}. 
Due to the fact that lepton PDFs peak at momentum fraction $ x=1 $, the inclusion of ISR effects is not expected to change the cross-sections dramatically. 
In a similar fashion to \cite{Bao:2022onq}, we estimate the magnitude of ISR by allowing for a single collinearly emitted photon in the final state with kinematic parameters such that the photon falls outside the acceptance of the detector. 
To this end, we enforce that the pseudorapidity $ \eta $ of the photon be in the domain $
2.5 < |\eta| < 1000 $ and require that its transverse momentum $p_T$
satisfy $ 0.001 \text{ GeV} < p_T < 0.1 \text{ GeV}$. 
The inclusive collinear photon is found to increase the cross-sections by $ \sim 30\% $ in the case of 10~TeV muon collisions. 
The cross-sections of figures \ref{fig:muon_collider_3} and \ref{fig:muon_collider_10} do not include the collinearly emitted photon, and one should take them as lower estimates of the real cross-section.

	Using the same discoverability criterion as before, we observe in figure~\ref{fig:muon_collider_3} that the cross-sections at the 3 TeV collider are large enough to explore  regions of the parameter space satisfying $ M_{Z'} \lesssim $ 5 TeV and $ m_\vartheta \lesssim 2.5 $ TeV with good sensitivity. 
Barring very small flavon charges $ q_{\theta} \ll 1 $, the 3~TeV collider would likely be able to discover or rule out flavonstrahlung in this region of $ m_{\vartheta} $ and $ M_{Z^\prime} $.
For $ Z^\prime $ masses larger than $ \sim 5 $ TeV, the amplitudes are increasingly suppressed by the off-shell $ Z' $ propagators, whereas flavon masses of order 3 TeV and greater are unreachable because the collider would be unable to produce an on-shell flavon in the final state.

Similarly, the 10~TeV muon collider has an excellent reach in the parameter space
region where $ M_{Z'} \lesssim 15 $ TeV and $ m_{\vartheta} \lesssim 8$ TeV. As
figure~\ref{fig:muon_collider_10} shows, the flavonstrahlung cross-sections are
very sensitive to the size of the coupling $ g_{Z^\prime} $ (they scale
as\footnote{Note that the cross-section scales as $g_{Z^\prime}^6 v_\vartheta^2$, but at a fixed value of
  $M_{Z^\prime}$, $v_\vartheta \propto 1/g_{Z^\prime}$.} $
  g_{Z^\prime}^4 $) compared with sensitivity to $m_\vartheta$ and $ M_{Z'} $. The cross-sections are greater
than around $ 10^{-1}$ fb for even the smallest allowed values of the coupling, meaning that
the 10~TeV collider would likely be able to cover all of the parameter
space , at least when $ q_{\theta} $ is of
order one or more.

\subsection{Summary of future collider prospects}
We see that the flavonstrahlung cross-sections are likely too small for the
process to be discovered at the HL-LHC, but a 100~TeV hadron collider and
a 3 or 10~TeV muon collider could have excellent discovery prospects
for multi-TeV mass flavons and $Z^\prime$ bosons.
At a qualitative level, this is very similar to the pure $Z^\prime$ search
prospects in the $B_3-L_2$ model, as shown in ref.~\cite{Azatov:2022itm}.

Comparing the 100~TeV hadron collider with the 10~TeV muon collider, we notice that for a given parameter space point, the flavonstrahlung cross-sections are two or three orders of magnitude greater at the muon collider. 
This is not surprising, considering that flavonstrahlung at a hadron collider occurs primarily through a $ bb $ partonic initial state, whereas the muon collider can harness almost the entire beam for the production of flavonstrahlung. 
The muon collider is limited by the kinematical threshold $ m_{\vartheta} \lesssim$ 8~TeV coming from the smaller beam energy, but even for the largest allowed couplings, the reach of the 100~TeV hadron collider is stopped in the same neighbourhood due to diminishing cross-sections.

To truly
estimate discovery prospects, one of course should calculate
backgrounds. However, it seems very likely that these can be controlled highly 
efficiently: with invariant mass cuts upon the reconstructed $Z^\prime$ from
its decay products and also from the reconstructed flavon particle, from its
decay products.

\section{Conclusions \label{sec:conc}}
The $B_3-L_2$ model is well motivated, being pertinent to the fermion mass
puzzle\footnote{In particular, the model describes why the CKM quark mixing
matrix elements $|V_{ub}| \ll 1$, $|V_{ts}|\ll 1$, $|V_{td}| \ll 1$ and
$|V_{cb}|\ll 1$.}~\cite{Alonso:2017uky,Bonilla:2017lsq,Allanach:2020kss} as well as
the \bsmm\ anomalies, 
and it is directly testable at
colliders.
Collider signatures remain currently 
relatively unstudied aside from a few $Z^\prime$ bump-hunts in di-fermion
invariant masses. Even these had to be reinterpreted from the experimental
papers because
most interpretations of various
bump-hunts in di-fermion invariant masses assumed family universality.
The $B_3-L_2$ model is far from the family universal limit;
for studies of $Z^\prime$ production in the $B_3-L_2$ model, see
refs.~\cite{Bonilla:2017lsq,Allanach:2020kss,Allanach:2021gmj,Azatov:2022itm}.

We have instead studied the prospects for flavonstrahlung, where the flavon
is produced along with
an associated $Z^\prime$. Our results indicate that the flavon will not be
directly discovered at the HL-LHC because its production cross-section is too
small, but in a future 3 or 10~TeV muon collider or a 100~TeV FCC, flavonstrahlung
discovery 
prospects are good. 
Flavons could also be produced at hadron colliders by traditional SM
Higgs production processes, provided that they mix with the Higgs (i.e.\
provided that
$\phi \neq 0$). Thus, they could be produced via 
gluon-gluon fusion, via weak boson fusion or via associated production with a
di-top fermion. The advantage of the flavonstrahlung process is that it is
present even in the limit of zero mixing with the SM Higgs field $\phi \rightarrow 0$. In fact, it becomes negligible
for maximal mixing with the SM Higgs field, but the model must be far from this 
limit because of current 
bounds from Higgs measurements, as shown in figure~\ref{fig:mixing_exclusion}.

Several other similar bottom-up models possess the flavonstrahlung signature,
for example the Third Family Hypercharge Model~\cite{Allanach:2018lvl} and its
variants~\cite{Allanach:2019iiy}, gauged muon minus tau lepton number~\cite{He:1990pn}
and several other gauged $U(1)$ family non-universal
models~\cite{Greljo:2021npi}. In the future, it may be of interest to compute
the cross-sections for these models too in order to see how they compare.

One may ask what the first hints would be in collider
data from beyond the SM effects of the $B_3-L_2$ model. The
answer, possibly, is the \bsmm\ anomalies that are currently being
investigated~\cite{Alonso:2017uky,Bonilla:2017lsq,Allanach:2020kss}. One can
also obtain deviations in $M_W$, as figure~\ref{fig:mixing_exclusion} 
implies. This figure also reminds us that (since we have applied the current
bounds, the sensitivity of which will increase with increased integrated
luminosity) the LHC may also observe deviations
from the SM limit in the signal strength of various Higgs 
cross-sections, should the flavon and Higgs fields mix. The first direct
evidence for the model, however, would
likely be from the classic $Z^\prime \rightarrow \mu^+ \mu^-$
resonance search, followed possibly by other $Z^\prime$ decay modes.
Flavon particle production (via the usual Higgs production modes, but
suppressed by 
powers of $\sin \phi$) could also be observed.
Flavonstrahlung may 
eventually be observed at future colliders,  with a resonant $\mu^+ \mu^-$
produced in association with a
flavon boson, which would primarily
decay to $W^+W^-$, $ZZ$ or $HH$, as 
displayed on figure~\ref{fig:flavon_BR}. 

\section*{Acknowledgements}
This work has been partially supported by STFC consolidated grant
ST/T000694/1. We thank the Cambridge Pheno Working Group for helpful
discussions. 

\bibliographystyle{JHEP-2}
\bibliography{pheno.bib}

\end{document}